\documentclass[twocolumn,showpacs,aps,floatfix,prd,nofootinbib]{revtex4}

\usepackage{graphicx}
\usepackage{dcolumn}
\usepackage{amsmath}
\usepackage{epsfig}
\usepackage{colordvi}
\usepackage{color}
\usepackage{hhline}

\RequirePackage{xspace}

\usepackage{relsize}
\def\babar{\mbox{\slshape B\kern-0.1em{\smaller A}\kern-0.1em
    B\kern-0.1em{\smaller A\kern-0.2em R}}}
     
\mathchardef\Upsilon="7107
\def\Y#1S{\ensuremath{\Upsilon{(#1S)}}\xspace}

\def\pep2{PEP-II}

\long\def\inst#1{\par\nobreak\kern 4pt\nobreak
  {\it #1}\par\vskip 10pt plus 3pt minus 3pt}
  
\begin{document}

\title{\large \bf
\boldmath
Study of $e^+e^-\to \pi^+\pi^-\pi^0$ process using initial state radiation
with \babar\
} 

\author{B.~Aubert}
\author{R.~Barate}
\author{D.~Boutigny}
\author{F.~Couderc}
\author{J.-M.~Gaillard}
\author{A.~Hicheur}
\author{Y.~Karyotakis}
\author{J.~P.~Lees}
\author{V.~Tisserand}
\author{A.~Zghiche}
\affiliation{Laboratoire de Physique des Particules, F-74941 Annecy-le-Vieux, France }
% ...
\collaboration{The \babar\ Collaboration}
\noaffiliation


\begin{abstract}
The process $e^+e^-\to \pi^+\pi^-\pi^0\gamma$ 
has been studied at a center-of-mass energy
near the \Y4S resonance 
using a 89.3 fb$^{-1}$ data sample collected with the \babar\ detector
at the \pep2 collider.  From the measured $3\pi$ mass
spectrum we have obtained the products of branching fractions for the $\omega$ and $\phi$ mesons, 
${\cal B}(\omega\to e^+e^-){\cal B}(\omega\to 3\pi)=(6.70\pm0.06\pm0.27)\times 10^{-5}$ and
${\cal B}(\phi\to e^+e^-){\cal B}(\phi\to 3\pi)=(4.30\pm0.08\pm0.21)\times 10^{-5}$,
and evaluated the
$e^+e^-\to \pi^+\pi^-\pi^0$ cross section for the
$e^+e^-$ center-of-mass energy range 1.05 to 3.00 GeV.
About 900 
$e^+e^-\to J/\psi\gamma\to \pi^+\pi^-\pi^0\gamma$ events have 
been selected and the branching fraction 
${\cal B}(J/\psi\to \pi^+\pi^-\pi^0)=(2.18\pm0.19)\%$ has been 
measured.
\end{abstract}

\pacs{13.66.Bc, 14.40.Cs, 13.25.Gv, 13.25.Jx, 13.20.Jf}

\maketitle


\setcounter{footnote}{0}

\section{ \boldmath Introduction}
\label{intro}
\begin{figure}
\includegraphics[width=.7\linewidth]{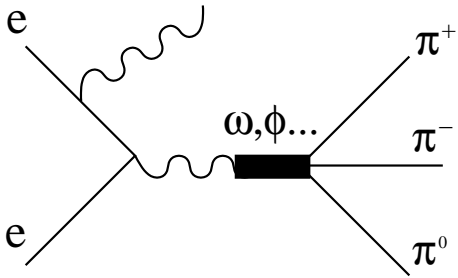}
\caption{The diagram for the $e^+e^-\to\pi^+\pi^-\pi^0\gamma$
process.}
\label{fei}
\end{figure}
The analysis of the process $e^+e^-\to \gamma + \mbox{hadrons}$,
where the photon emission is caused by initial state radiation (ISR),
can be used to measure the $e^+ e^-$ annihilation cross section into hadrons
over a wide range of center-of-mass (c.m.) energy in a single
experiment~\cite{arbus,kuhn,ivanch}.
In these events, the invariant mass, $\sqrt{s^\prime}$,
of the final state hadronic system corresponds to the ``effective''
center-of-mass energy after ISR.

This method, currently used both at the KLOE experiment at DAFNE
(Frascati)~\cite{KLOE}
and at the \babar\ experiment at the \pep2\
$B$ Factory (SLAC), is 
applied here to study
the process $e^+ e^- \rightarrow \pi^+ \pi^- \pi^0$ at low energies (1
to 3 GeV).

The Born cross section for the $e^+e^-\to \mbox{hadrons}+\gamma$ process
(Fig.~\ref{fei})
integrated over the momenta of the hadrons is given by 
\begin{equation}
\frac{{\rm d}\sigma(s,x,\theta)}{{\rm d}x\,{\rm d}\cos{\theta}} = W(s,x,\theta)\, \sigma_0(s(1-x)),
\label{eq1}
\end{equation}
where $\sqrt{s}$ is the $e^+e^-$ center-of-mass energy,
$x\equiv{2E_{\gamma}}/{\sqrt{s}}$,
$E_{\gamma}$ and $\theta$ are the photon energy and polar angle in 
the c.m. frame, and $s(1-x)=s^\prime$, already mentioned above.
Here $\sigma_0$ is defined as 
the Born cross section for $e^+e^-\to \mbox{hadrons}$.
The so-called radiator function (see, for example, Ref.~\cite{ivanch})
\begin{equation}
W(s,x,\theta)=\frac{\alpha}{\pi x}\left(\frac{2-2x+x^2}{\sin^2\theta}-
\frac{x^2}{2}\right)
\label{eq2} 
\end{equation}
describes the probability of ISR photon emission for $\theta\gg m_e/\sqrt{s}$.
Here $\alpha$ is the fine structure constant and $m_e$ is the electron mass. 
The ISR photons are emitted predominantly at small angles relative to
the initial electron or positron directions; however about 10\% of the photons
have c.m. polar angles in the range $30^\circ<\theta<150^\circ$.
In the present analysis, we require that the ISR photon is detected.

The  differential cross section for ISR production
of a narrow resonance (vector meson $V$),
such as $J/\psi$, decaying into the final state $f$ can be calculated 
using~\cite{ivanch}
\begin{equation}
\frac{{\rm d}\sigma(s,\theta)}{{\rm d}\cos{\theta}}
=\frac{12\pi^2 \Gamma(V\to e^+e^-) {\cal B}(V\to f)}{m\, s}\,
W(s,x_0,\theta),
\label{eq3}
\end{equation}
where $m$ and $\Gamma(V\to e^+e^-)$ are the mass and electronic
width of the vector meson $V$, $x_0 = 1-{m^2}/{s}$,
and ${\cal B}(V\to f)$ is the branching fraction of $V$
into the final state $f$.
Therefore, the measurement of the number of $J/\psi \to 3\pi$
decays
in $e^+ e^- \to 3\pi\gamma$ determines the product of the
electronic width and the branching fraction:
$\Gamma(J/\psi \to e^+e^-){\cal B}(J/\psi \to 3\pi)$.

The $e^+e^-\to \pi^+\pi^-\pi^0$ cross section in the energy region
$\sqrt{s^\prime}\lesssim 1$ GeV is dominated by the 
$\omega(782)$ and $\phi(1020)$ 
mesons\footnote{Throughout this paper, $2\pi$, $3\pi$, and $4\pi$
mean  $\pi^+\pi^-$, $\pi^+\pi^-\pi^0$, and $\pi^+\pi^-\pi^0\pi^0$,
respectively. We also use the notations $\omega$, $\phi$,
$\omega^\prime$, and $\omega^{\prime\prime}$ for $\omega(782)$,
$\phi(1020)$, $\omega(1420)$, and $\omega(1650)$.}.
This energy region has been studied in many 
experiments with high statistics and the Particle Data Group
(PDG) parameters~\cite{pdg} for
$\omega$ and $\phi$ mesons have relatively high precision
(2--3\% for ${\cal B}(V\to e^+e^-){\cal B}(V\to 3\pi)$ and about 1\% for
the total widths).

The energy region above the $\phi$ was studied in two experiments:
SND~\cite{SND2002} for energies up to 1.4 GeV with statistical precision
about 10\% and DM2~\cite{DM2} for energies in the 1.34--2.40 GeV range with
statistical precision about 25\%.
As pointed out in Ref.~\cite{SND2002},
there is a significant systematic shift between these two
datasets, and the DM2 data need to be scaled by
a factor $1.72\pm0.24$ in order to fit with those of SND.

In this energy region, the $e^+ e^- \to 3\pi$ cross section
is generally  described as the sum of two resonances
$\omega^\prime(1420)$ and
$\omega^{\prime\prime}(1650)$.
So cross section measurement allows the determination of the $\omega^\prime$
and $\omega^{\prime\prime}$ parameters.
Masses, widths, and decay modes for these resonances are not well 
established. The PDG~\cite{pdg} gives only estimates for these 
parameters.

The main goal of this analysis is an independent measurement of
the $e^+e^-\to \pi^+\pi^-\pi^0$ cross section in the energy region
from 1.05 to 3.00 GeV. The aim is to significantly improve the precision of
the cross section for energies above 1.4 GeV. Our data in the
$\omega-\phi$ region can be compared with the more precise $e^+e^-$ data
in this region and the difference can be used 
to check our systematic error estimation.

We study $J/\psi$ production in the process
$e^+e^-\to \pi^+\pi^-\pi^0\gamma$, 
and measure the product
$\Gamma(J/\psi\to e^+e^-) {\cal B}(J/\psi\to 3\pi)$.
The branching fraction ${\cal B}(J/\psi\to 3\pi)$ is then
determined using the known value of
$\Gamma(J/\psi\to e^+e^-)$~\cite{BAD602}.
The decay $J/\psi\to 3\pi$ has been studied in many experiments
\cite{JEAN,BARTEL,BRANDELIK,ALEXANDER,FRANKLIN,COFFMAN,BAI} but only three
of them measured the decay rate without any restrictions on the invariant mass
of the two-pion system. Three results for ${\cal B}(J/\psi\to 3\pi)$ are
$(1.6\pm0.4)\%$~\cite{ALEXANDER}, $(1.5\pm0.2)\%$~\cite{FRANKLIN},
$(1.42\pm0.19)\%$~\cite{COFFMAN}. The average of these measurements is 
$(1.47\pm0.13)\%$. This value is in significant disagreement
with more recent result from the BES Collaboration~\cite{BES3pi}:
$(2.10\pm0.12)\%$.

\section{ \boldmath The \babar\ detector and data samples}
\label{detector}
In this paper a data sample of 89.3 fb$^{-1}$, collected by the 
\babar\ detector~\cite{ref:babar-nim} at the \pep2\ asymmetric-energy storage
ring, is analyzed. At \pep2, 9-GeV electrons collide with 3.1-GeV positrons
at a center-of-mass energy of 10.6 GeV ($\Upsilon$(4S) resonance).

Charged-particle tracking for the BABAR detector is
provided by a five-layer silicon vertex tracker (SVT) and
a 40-layer drift chamber (DCH), operating in a 1.5-T axial
magnetic field. The transverse momentum resolution
is 0.47\% at 1 GeV/$c$. Energies of photons and electrons
are measured with a CsI(Tl) electromagnetic calorimeter
(EMC) with a resolution of 3\% at 1 GeV. Charged-particle
identification is provided by ionization measurements in
the SVT and DCH, and by an internally reflecting ring-imaging
Cherenkov detector (DIRC). Muons are identified
in the solenoid's instrumented flux return,
which consists of iron plates interleaved with resistive
plate chambers.

Signal and background ISR processes are simulated using Monte
Carlo (MC) event generators based on the computer code 
described in Ref.~\cite{ckhhad}.
The event generator for the $e^+e^-\to 3\pi\gamma$ reaction uses a
model with an intermediate $\rho\pi$ state.
The background process $e^+e^-\to \pi^+\pi^-\pi^0\pi^0\gamma$
is simulated with $\omega\pi^0$ and $a_1(1260)\pi$ intermediate
states in the proportion that matches existing experimental data.
The extra soft-photon radiation is generated with the use of the structure
function method of Ref.~\cite{strfun} and the PHOTOS package~\cite{PHOTOS} for
electron and charged hadron bremsstrahlung, respectively.
Since the polar-angle distribution of the
ISR photon is peaked near $0^\circ$ and $180^\circ$, the events
are generated with the restriction on the photon polar angle in the c.m. frame,
$20^\circ<\theta<160^\circ$. We also require that
the invariant mass of the hadron system and ISR photon together is greater than
8 GeV/$c^2$. This second cut restricts the maximum energy of extra photons
emitted by the initial particles. 
The background processes
$e^+e^-\to \pi^+\pi^-\gamma,\:\mu^+\mu^-\gamma$
are generated with the Phokhara program~\cite{Phokhara}, which includes
next-to-leading-order QED corrections and simulates the emission of
two hard photons at large angle by the initial particles.
The background from $e^+e^-\to q\bar{q}$ and $e^+e^-\to \tau^+\tau^-$ is 
simulated with JETSET~\cite{Jetset} and KORALB~\cite{KORALB} packages, 
respectively.
The interaction of the generated particles with the \babar\ detector 
and the detector response are simulated using the GEANT4~\cite{ref:geant4} 
package. The simulation takes into account the variation of the detector and 
accelerator conditions, and in particular describes the beam-induced background, 
which leads to the appearance of photons and tracks overlapping on the events 
of interest.

\section{ \boldmath Event selection}
\label{selection}
The initial selection of $e^+e^-\to \pi^+\pi^-\pi^0\gamma$
candidates requires that
all the final particles are detected inside a fiducial volume.
(All kinematic variables used in the paper are defined in the laboratory
frame unless otherwise stated.)
Since a significant fraction of the events contain beam-generated
spurious tracks and photons,
we select events with two or three tracks and
at least three photons that have energies above 100 MeV and polar angles in
the range $23^\circ < \theta < 137.5^\circ$ (the corresponding angular range in the
c.m. frame is $38^\circ<\theta<154^\circ$).
One of the photons is required to have an energy in the c.m. frame above 3 GeV.
Two of the tracks must originate from the interaction point, have
a transverse momentum above 100 MeV/$c$ and be in the
polar angle region between $23^\circ$ and $140^\circ$.
Background events from the process $e^+e^-\to e^+e^-\gamma$
are suppressed by requiring the ratio of the calorimeter-deposited
energy to the track momentum, $E_{EMC}/p$, to be below 0.9 
for the two highest-momentum tracks.

\begin{figure}
\includegraphics[width=0.9\linewidth]{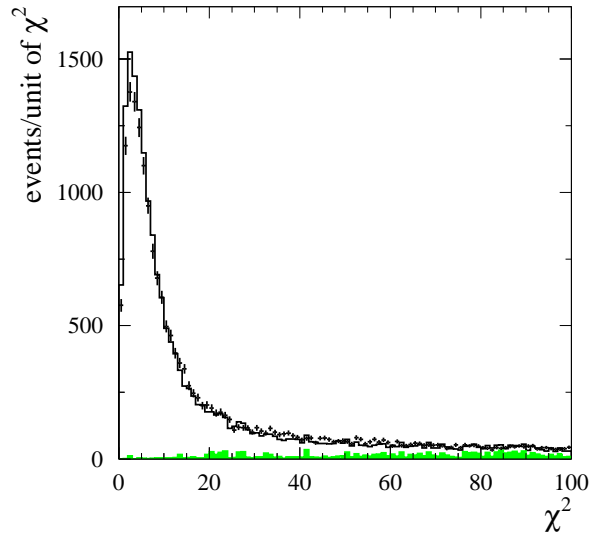}
\caption{The $\chi^2$ distributions for data (points with error bars) and 
simulated (histogram) events from the $\omega$ mass region. The shaded histogram 
shows the distribution for simulated background events.}
\label{chi_omega}
\end{figure}
\begin{figure}
\includegraphics[width=0.9\linewidth]{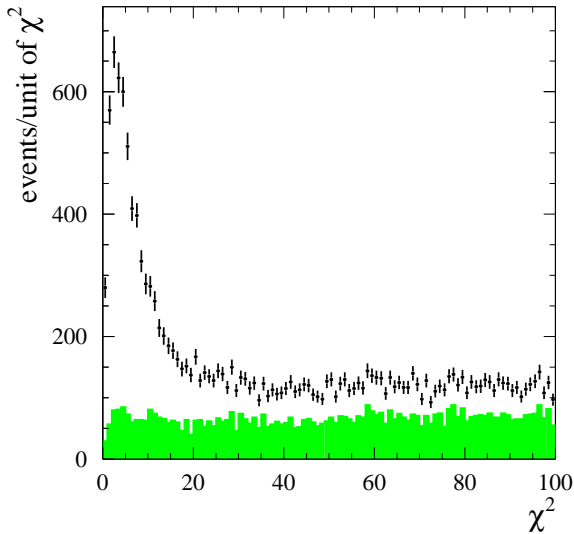}
\caption{The $\chi^2$ distributions for data
from the mass range $1.05<M_{3\pi}<3.00$ GeV/$c^2$.
The shaded histogram shows events rejected
by background suppression cuts.}
\label{chi_13}
\end{figure}
The photon with greatest c.m. energy is assumed to be the ISR
photon. The remaining photons are paired to form candidate $\pi^0$s,
requiring that their invariant mass must be in the range 0.07 to 0.20
GeV/$c^2$. A kinematic fit is applied to the selected event, imposing
energy and momentum conservation, and constraining the candidate $\pi^0$
invariant mass. The MC simulation does not accurately
reproduce the shape of the resolution function for the photon
energy. This leads to a difference in the distributions of the
$\chi^2$ of the kinematic fit for data and simulation. To reduce this
difference only the measured direction of the ISR photon is
used in the fit; its energy is a free fit parameter.
In the case of events with three tracks, the fit uses the parameters of the
two tracks with opposite charge that have the minimum distance from
the interaction point in the azimuthal plane.
For events with more than three photons all possible combinations
of photons are tested and the one with minimum $\chi^2$ is used.
The events with very high $\chi^2$ ($>10000$) are considered as
not reconstructed.

The $\chi^2$ of the kinematic fit is used to discriminate real 
$3\pi\gamma$ events from background.
Figure~\ref{chi_omega} shows the $\chi^2$
distribution for events from the $3\pi$ mass region near
the $\omega$ mass (0.75--0.82 GeV/$c^2$),
where the contribution of background processes is small.
Events with $\chi^2 < 40$ are selected to analyze the $3\pi$ mass spectrum.
The rest of the sample ($10000 > \chi^2 > 40$) is used to study both background
processes and possible selection-efficiency  corrections due to
data-MC simulation differences in the $\chi^2$ distribution.
The $\chi^2$ distribution
for masses $1.05<M_{3\pi}<3.00$ GeV/$c^2$ (Fig.~\ref{chi_13}) shows
that a significant fraction of events in this mass range 
correspond to background processes.

The main sources of background are other ISR processes
($e^+e^-\to\pi^+\pi^-\pi^0\pi^0\gamma,\: \pi^+\pi^-\gamma,\:
K^+K^-\pi^0\gamma$, etc.) and $e^+e^-$ annihilation to $q\bar{q}$ 
and $\tau^+\tau^-$. Additional background suppression cuts are 
applied to improve the signal-to-background ratio in the mass region of 
interest.

The ISR events with kaons in the final state
($e^+e^-\to K^+K^-\pi^0\gamma$, $e^+e^-\to K^+K^-\gamma$)
are suppressed using the kaon identification based
on $dE/dx$ measurements in the tracking devices, and the value of the
Cherenkov angle and the number of photons measured in the DIRC.
The requirement that none of the charged tracks is identified as a kaon
rejects about 95\% of the kaon-induced background
with only 4\% loss of $3\pi\gamma$ events.

The radiative events $e^+e^-\to\pi^+\pi^-\gamma$ and
$e^+e^-\to\mu^+\mu^-\gamma$ with
extra photons having $\gamma\gamma$ invariant mass close to that of a $\pi^0$
are suppressed by a cut on the $\pi^0$ energy. The cut $E_{\pi^0}>0.4$ GeV
rejects 80\% of $e^+e^-\to \pi^+\pi^-\gamma$ and $e^+e^-\to\mu^+\mu^-\gamma$
events with about 4\% loss of $3\pi\gamma$ events.

The process $e^+e^-\to\pi^+\pi^-\pi^0\pi^0\gamma$ with a soft $\pi^0$
is the main source of background for the process under study. Some
fraction of $4\pi\gamma$ events reconstructed as $\pi^+\pi^-\pi^0\gamma$
contain a $\pi^0$ among extra
photons.  We select these events by performing a kinematic fit for the
$4\pi\gamma$ hypothesis. The requirement $\chi^2_{4\pi\gamma}>40$ rejects
about 40\% of $4\pi\gamma$ events and only 2\% of $3\pi\gamma$ events. 

The main source of background from $e^+e^-\to\tau^+\tau^-$ is
the events in which both $\tau$'s decay into $2\pi\nu$. The hard photon arises
from a $\pi^0$ decay. Since the $\tau \to 2\pi\nu$ decay proceeds mainly via
$\rho\nu$, such events must have the invariant mass of the most energetic
photon and one of the charged pions peaked near the $\rho$ mass.
The cut $M_{\pi\gamma}>1.5$ GeV/$c^2$ almost fully rejects
$e^+e^-\to\tau^+\tau^-$ background and leads to only 0.3\% loss of
$3\pi\gamma$ events. The remaining $\tau^+\tau^-$ background
is estimated to be less than 0.1\% of $3\pi\gamma$ events.

Another possible background source is events from $e^+e^-$
annihilation into hadrons containing a very energetic $\pi^0$.
A fraction of these events can
be seen in the distribution of invariant mass ($M_{\gamma\gamma}^\ast$)
of two photons, one of which is the most energetic in an event.
The second photon in the pair is required to have an energy above 100 MeV.
Once all possible photon pair combinations are checked, the one with closest
invariant mass to the $\pi^0$ mass is chosen.
The events with
$0.10<M_{\gamma\gamma}^\ast<0.17$ GeV/$c^2$ are rejected. 

The $\chi^2$ distribution for all rejected events is shown 
as the shaded histogram in Fig.~\ref{chi_13}.
It is seen from Fig.~\ref{chi_13} that we reject more than 60\% of background
events without significant loss of signal events.
\section{ \boldmath Background calculation and subtraction}
\label{subtraction}
\begin{figure}
\includegraphics[width=0.9\linewidth]{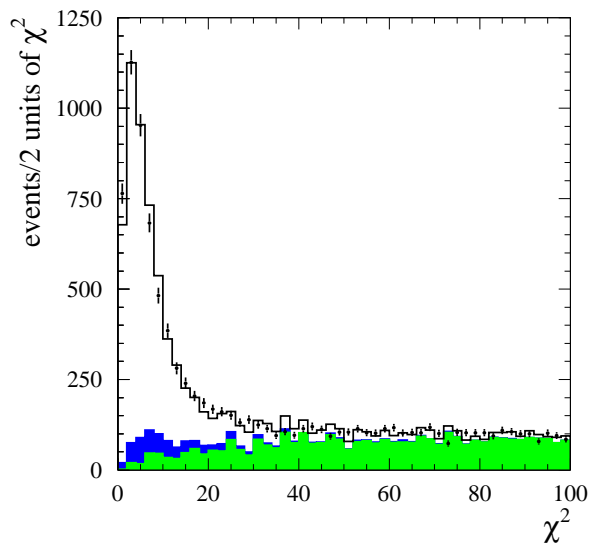}
\caption{The $\chi^2$ distribution for events from the mass 
range $1.05<M_{3\pi}<3.00$ GeV/$c^2$ after background suppression cuts.
The points with error bars show the data distribution. The histogram is the
sum of simulated distributions 
for $e^+e^-\to\pi^+\pi^-\pi^0\gamma$ and background processes.
The dark and lightly shaded histograms show the distributions for
$e^+e^-\to\pi^+\pi^-\pi^0\pi^0$ and other background processes, respectively.} 
\label{chi2_bkg_calc}
\end{figure}

\begin{figure}
\includegraphics[width=0.9\linewidth]{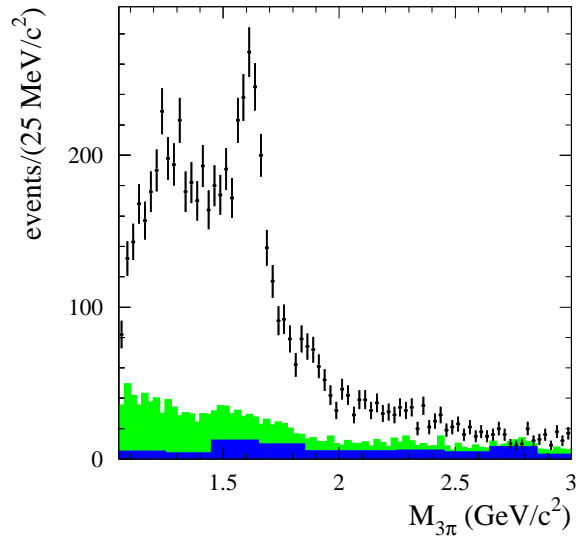}
\caption{The $3\pi$ mass spectrum for data events with $\chi^2<40$.
The lightly shaded histogram shows the mass spectrum for 
calculated background.
The dark histogram is $e^+e^-\to\pi^+\pi^-\pi^0\pi^0$ background.}
\label{upperphi_m}
\end{figure}

The $\chi^2$ distribution and $3\pi$ mass spectrum for
data and for simulation of $e^+e^-\to\pi^+\pi^-\pi^0\gamma$ and background
processes  after imposing the background suppression cuts are shown
in Figs.~\ref{chi2_bkg_calc} and~\ref{upperphi_m}.
The remaining background can be divided into two classes with
different $\chi^2$ distributions. The first class includes
$e^+e^-\to K^+K^-\pi^0 \gamma$ and $e^+e^-\to \pi^+\pi^-\pi^0\pi^0$
processes, which have $\chi^2$ distributions peaked at low $\chi^2$.
The second class includes all other background processes.
The simulated $\chi^2$ distribution for these processes is shown by the
lightly shaded histogram in Fig.~\ref{chi2_bkg_calc}.
The background events from the 
first class must be subtracted bin-by-bin from the $M_{3\pi}$ spectrum. 

The mass distribution for kaon events surviving the selection ($N_{0K}$)
can be obtained from the distribution of events with two identified kaons
($N_{2K}$):
$N_{0K}(M_{3\pi})=N_{2K}(M_{3\pi})R_K.$
The coefficient $R_K$ is determined from the simulation, which uses 
kaon identification efficiency corrections
obtained from a pure kaon sample from $D^\ast$ decays
to improve the agreement of data and simulation. 
The value of $R_K$ is found to 
be independent of $M_{3\pi}$ and equal to $0.09\pm0.01$. 
The total fraction of kaon events in the final event sample is estimated
to be 0.4\%.

The simulation shows that about 80\% of the $e^+e^-\to q\bar{q}$ events
that pass the selection criteria have $\pi^+\pi^-\pi^0\pi^0$ as the final 
state.
In order to cross-check the value of the yield given by JETSET
fragmentation for this particular final
state, we use the following procedure to extract the mass distribution for
$e^+e^-\to \pi^+\pi^-\pi^0\pi^0$ from experimental data.
We select
events with two charged particles and four photons with energy more than
0.1 GeV, at least one of them with c.m. energy more than 3 GeV, perform
a kinematic fit to the  $e^+e^-\to\pi^+\pi^-\pi^0\pi^0$ hypothesis
and require the $\chi^2$ of this fit to be less than 20.
The number of selected $4\pi$ events is found to be about 15\% less than
the number expected from JETSET. We also studied various two- and
three-particle mass distributions and find that both in data and
simulation the $e^+e^-\to\pi^+\pi^-\pi^0\pi^0$ process proceeds via the
$\rho\pi\pi$ state. No other intermediate states are seen.
To obtain the $3\pi$ mass spectrum for $e^+e^-\to\pi^+\pi^-\pi^0\pi^0$
events reconstructed in the $e^+e^-\to\pi^+\pi^-\pi^0\gamma$ hypothesis
($({\rm d}N/{\rm d}m)_{3\pi\gamma}$), we multiply the $3\pi$ mass spectrum of
selected $4\pi$ events ($({\rm d}N/{\rm d}m)_{4\pi}^{exp}$) by
the ratio of corresponding simulated distributions:
$({\rm d}N/{\rm d}m)_{3\pi\gamma}^{MC}/({\rm d}N/{\rm d}m)_{4\pi}^{MC}$. 
The resulting mass
spectrum is shown in Fig.~\ref{upperphi_m} as the dark shaded histogram.
The fraction of this background does not
exceed 10\% in the mass region below 2 GeV/$c^2$ and rises at 
higher masses.

\begin{figure*}
\begin{minipage}[t]{0.47\textwidth}
\includegraphics[width=\textwidth]{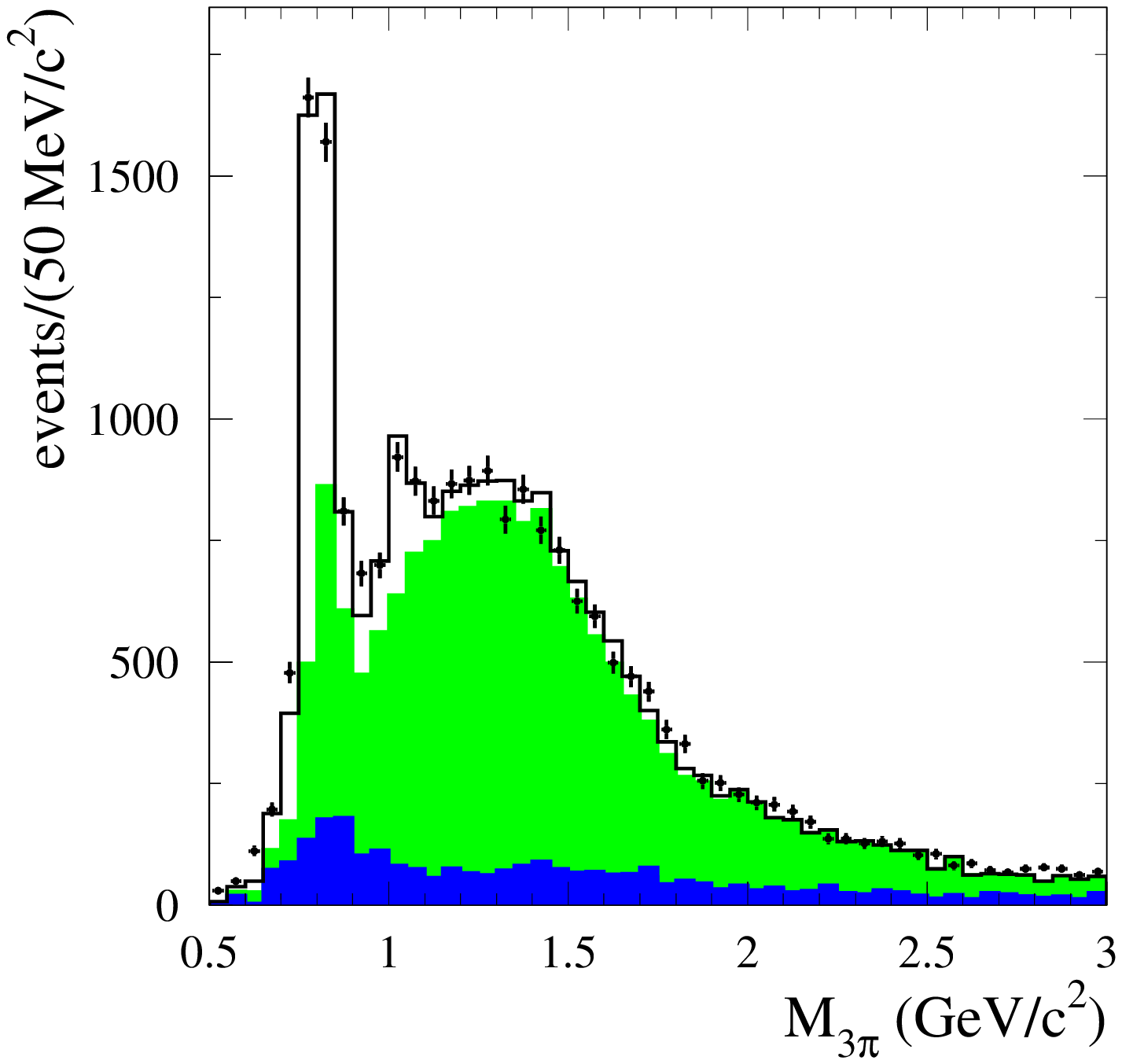}
\caption{The $M_{3\pi}$ spectrum for events with
$100<\chi^2<500$ and $E_{\pi^0}>1$ GeV.
The points with error bars are data.
The histogram is the sum of spectra for simulated events of
$e^+e^-\to\pi^+\pi^-\pi^0\gamma$ and background processes.
The lightly shaded histogram is the sum of all background processes.
The difference between the dark and lightly shaded histograms shows the
$e^+e^-\to\pi^+\pi^-\pi^0\pi^0\gamma$ contribution.}
\label{mass_4p}
\end{minipage}
\hfill
\begin{minipage}[t]{0.47\textwidth}
\includegraphics[width=\textwidth]{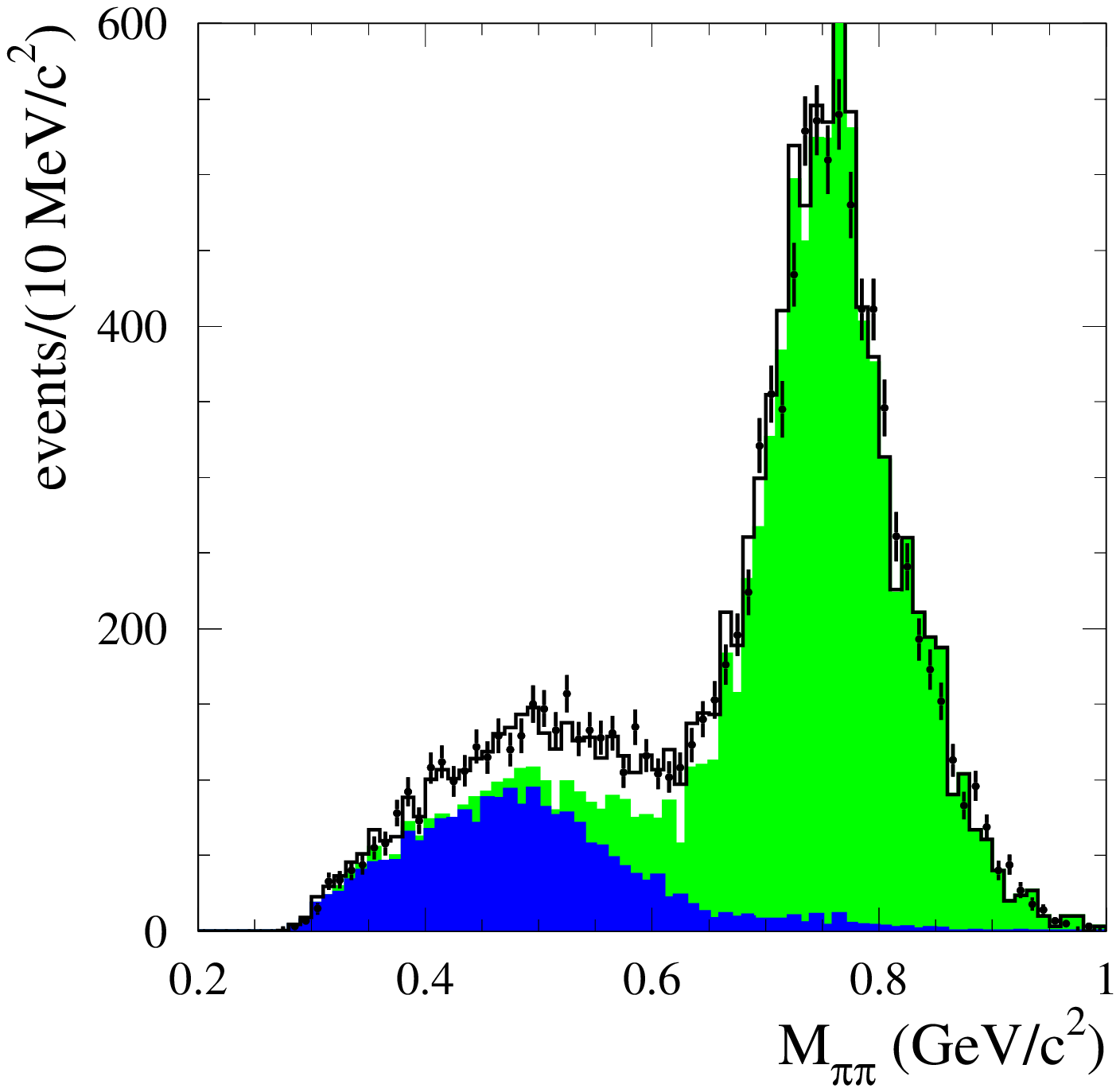}
\caption{$\pi^+\pi^-$ invariant mass spectrum for events with
$100<\chi^2<500$ and $0.85<M_{3\pi}<1.05$ GeV/$c^2$.
The points with error bars are data.
The histogram is the sum of spectra for simulated events of
$e^+e^-\to\pi^+\pi^-\pi^0\gamma$ and background processes.
The lightly shaded histogram is the sum of all background processes.
The difference between the dark and lightly shaded histogram shows
the $e^+e^-\to\pi^+\pi^-\gamma$ contribution.}
\label{rhopp1}
\end{minipage}
\end{figure*}
The main contribution to the second class of background
events comes from the $e^+e^-\to\pi^+\pi^-\pi^0\pi^0\gamma$ and
$e^+e^-\to\pi^+\pi^-\gamma$ processes.
To estimate the accuracy of the MC simulation prediction of the background
level for these processes we use
events with $\chi^2>40$. Fitting distributions of the $3\pi$
and $\pi^+\pi^-$ invariant masses for these predominantly background events
with a sum of distributions for the
signal and background processes, we find scale factors of $(1.00\pm 0.25)$ for
$4\pi\gamma$ and $(3\pm1)$ for $2\pi\gamma$.
Figures~\ref{mass_4p} and~\ref{rhopp1} show the fitted
$M_{3\pi}$ and $M_{\pi^+\pi^-}$ distributions for events with
$100<\chi^2<500$.
The quoted errors in the scale factors are much larger than the statistical
errors from the fits and originate from systematic effects. For
$4\pi\gamma$ events the error is determined by the dependence of the
scale factor value on $\chi^2$. For $2\pi\gamma$ events we observe a
significant difference in the shape of the $3\pi$ mass distribution
between data and simulation and the 30\% systematic error has been
assigned to cover the uncertainty on the mass spectrum.

We attribute this difference and the large value
of the $2\pi\gamma$ scale factor to the inaccuracy of the
simulation of $\pi$-meson nuclear interactions in the EMC. This leads
to a difference between data and MC simulation in the multiplicities of 
spurious photons arising from the nuclear interaction, and
their energy and angular distributions. This conclusion is supported
by the fact that the scale factor for $e^+e^-\to\mu^+\mu^-\gamma$
background events is found to be close to unity. The 
$e^+e^-\to\mu^+\mu^-\gamma$ events were selected using muon identification
criteria based on EMC and IFR information. The level of 
$e^+e^-\to\mu^+\mu^-\gamma$ background is found to be (5--15)\% of 
the $2\pi\gamma$ background.

Another source of the second class of background, non-$4\pi$
$e^+e^-\to q\bar{q},\;q=u,d,s,c$, does not exceed 10\% of the total
level of background for masses up to 2.5 GeV/$c^2$. 
Since the accuracy of the JETSET prediction for the
$e^+e^- \to \pi^+\pi^-\pi^0\pi^0$ process is at the 15\% level,
we conservatively assume that the predictions for non-$4\pi$ backgrounds
are accurate to better than 50\%.

The production of a $\pi^0$ or a photon with c.m.
energy more than 3 GeV in $B\bar{B}$ events is kinematically forbidden.
We therefore do not expect any background from $B$ decays. 

The calculated level of background is about 0.6\% in the $\omega$ mass region
and 1.4\% in the $\phi$ mass region. The $3\pi$ mass
spectrum above the $\phi$ for data and background events is shown in
Fig.~\ref{upperphi_m}. The level of background is 15\% at 1.5 GeV/$c^2$
and 50\% at 3 GeV/$c^2$. The systematic uncertainty on the background
level is about 25\% below 2 GeV/$c^2$. For higher masses
the fractional uncertainty grows
due to an uncertainty in the $q\bar{q}$ background calculation,
a model dependence in the simulation of $e^+e^-\to 2\pi\gamma$ and
$e^+e^-\to 4\pi\gamma$
(the processes $e^+e^-\to 2\pi$ and $e^+e^-\to 4\pi$
have not been measured for $e^+e^-$ c.m. energies above 2 GeV), and 
the possible contribution of unaccounted
ISR processes ($e^+e^-\to \pi^+\pi^-\, 3\pi^0\gamma,\; \pi^+\pi^-\, 4\pi^0\gamma$, etc.).

We therefore use two different methods for background subtraction. For masses
below 1.05 GeV/$c^2$, where the level of background is low, we subtract 
the calculated 
background. For higher masses we use the procedure of statistical subtraction 
based on the difference in $\chi^2$ distributions of signal and background events.

The statistical subtraction procedure is as follows. For each
mass bin we find the numbers of events with $\chi^2\leq 20$ ($N_1$) and
$20<\chi^2<40$ ($N_2$) and calculate the number of signal and
background events with $\chi^2<40$ as
\begin{equation}
N_s=\frac{(1-\beta)N_1-\beta N_2}{\alpha-\beta},
\label{bsubeq1}
\end{equation}
\begin{equation}
N_{bkg}=\frac{\alpha N_2-(1-\alpha) N_1}{\alpha-\beta},
\label{bsubeq2}
\end{equation}
where $\alpha,\beta=N_1/(N_1+N_2)$ for pure signal and
background events, respectively. $N_1$ and $N_2$ in 
Eq.~(\ref{bsubeq1}--\ref{bsubeq2})
do not contain events $e^+e^-\to K^+K^-\pi^0\gamma$ and 
$e^+e^-\to\pi^+\pi^-\pi^0\pi^0$
processes, which are subtracted from both mass distributions bin-by-bin.

\begin{table*}[tb]
\caption{The values of $\beta=N(\chi^2<20)/N(\chi^2<40)$ calculated for
different background processes in four mass regions.
$w$ is the relative contribution of the process to the total background.
On the average, each value of $\beta$ is weighted by the corresponding value 
of $w$.}
\begin{ruledtabular}
\begin{tabular}{lcccccccccc} 
&\multicolumn{2}{c}{$1.05<M_{3\pi}\leq 1.40$}
&\multicolumn{2}{c}{$1.4<M_{3\pi}\leq 2.0$}
&\multicolumn{2}{c}{$2.0<M_{3\pi}\leq 2.5$}
&\multicolumn{2}{c}{$2.5<M_{3\pi}<3.0$}
&\multicolumn{2}{c}{$1.05<M_{3\pi}<3.00$} \\
process                                      &$w$(\%)&  $\beta$      &$w$(\%)&  $\beta$      &$w$(\%)&  $\beta$      &$w$(\%)&  $\beta$      &$w$(\%)&   $\beta$   \\
\hline
$\pi^+\pi^-\pi^0\pi^0\gamma$          & 68  & $0.30\pm0.02$ & 70  & $0.33\pm0.02$ & 67  & $0.34\pm0.03$ & 35  & $0.36\pm0.05$ & 66  & $0.32\pm0.01$\\
$\pi^+\pi^-\gamma$, $\mu^+\mu^-\gamma$& 28  & $0.33\pm0.10$ & 20  & $0.35\pm0.14$ & 27  & $0.54\pm0.20$ & 42  & $0.29\pm0.19$ & 26  & $0.36\pm0.07$\\
$q\bar{q}$, $\tau^+\tau^-$            &  4  & $0.39\pm0.12$ & 10  & $0.40\pm0.08$ &  6  & $0.50\pm0.20$ & 23  & $0.30\pm0.11$ &  8  & $0.38\pm0.06$\\
\hline
Average                               &     & $0.32\pm0.03$ &     & $0.34\pm0.03$ &     & $0.40\pm0.07$ &     & $0.32\pm0.09$ &     & $0.33\pm0.02$\\
\end{tabular}
\end{ruledtabular}
\label{beta}
\end{table*}
The coefficient $\beta$ is determined from the simulation.
Its values for the main background processes
in four mass regions are listed in Table~\ref{beta}. It is seen that there is
no significant dependence of $\beta$ on the mass and that the three main
background process have consistent values of $\beta$.
We therefore use the average value $\beta=0.33\pm0.02\pm0.05$.
The variation of the $\beta$ values for different processes was used
as an estimate of the systematic error.

The values of $\alpha$ at the $\phi$ and $J/\psi$ masses are extracted from
data. In the $\phi$ mass region we determine the ratio $N_1/(N_1+N_2)$
for pure signal events from
the experimental $\chi^2$ distribution with subtracted background.
The value of $\alpha$ at the $J/\psi$ mass is measured by another method.
The numbers of $J/\psi$ events for different $\chi^2$ cuts are determined
using a mass-sideband subtraction method (see Sec.~\ref{jpsi}). The resulting
values of $\alpha$ are
$0.879\pm 0.006\pm 0.005$ at the $\phi$ mass
and $0.882\pm 0.014$ at the $J/\psi$ mass.
The systematic error on $\alpha$ at the $\phi$
mass is estimated by varying the calculated background level by $\pm$25\%.
For the mass range 1.05-3.00 GeV/$c^2$
we use a linear interpolation between the $\phi$ and
$J/\psi$ values of $\alpha$.

\begin{figure}
\includegraphics[width=0.9\linewidth]{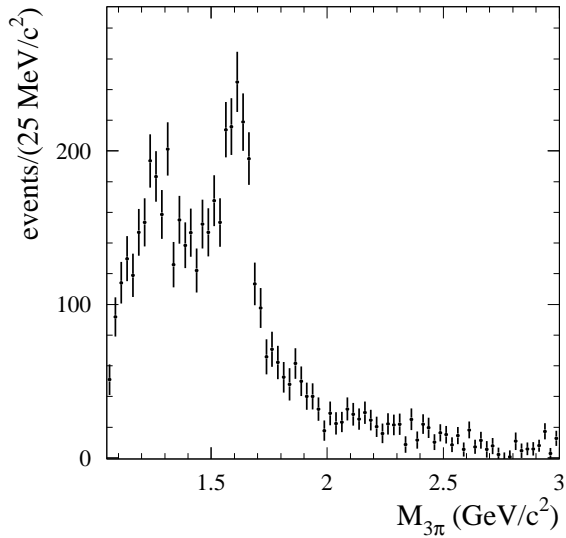}
\caption{The $3\pi$ mass spectrum for data obtained after the 
statistical background-subtraction procedure.}
\label{signal-1-3}
\end{figure}
\begin{figure}
\includegraphics[width=0.9\linewidth]{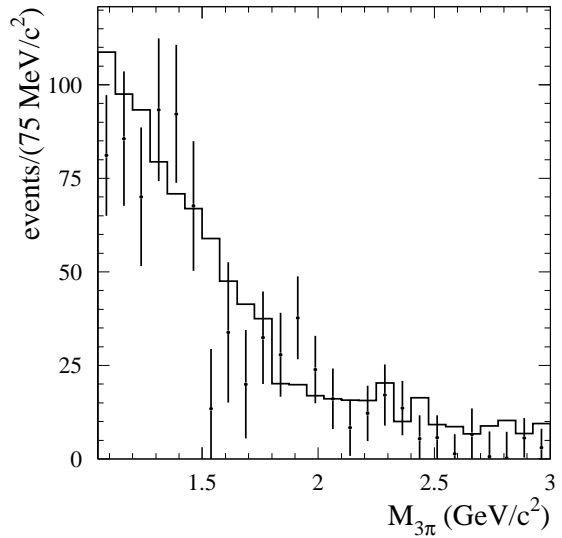}
\caption{The $3\pi$ mass spectrum for background events 
obtained from the statistical background-subtraction 
procedure (points with error bars). 
The histogram shows
the background mass distribution expected from MC simulation.}
\label{bkg-1-3}
\end{figure}
The $3\pi$ mass spectrum obtained after this statistical background subtraction
is shown in Fig.~\ref{signal-1-3}. The $3\pi$ mass distribution for the 
background events obtained using Eq.~(\ref{bsubeq2}) is shown in 
Fig.~\ref{bkg-1-3} and compared with the MC simulation.
The simulation describes well both the shape of the 
background spectrum and the total number of events up to at least 
2.5 GeV/$c^2$. The shapes of the signal and background distributions are quite
different.

\section{Detection efficiency}\label{sdetef}
\begin{figure}
\includegraphics[width=0.9\linewidth]{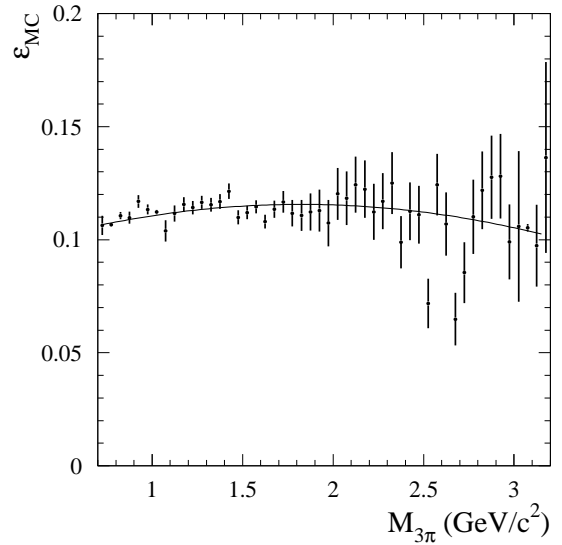}
\caption{The $3\pi$-mass dependence of the detection efficiency obtained
from MC simulation. The line is the fit to a second-order polynomial.} 
\label{deteff}
\end{figure}

To first approximation the detection efficiency is determined from MC
simulation as the ratio of the true $3\pi$ mass distribution computed
after and before applying the selection criteria.
The detection efficiency calculated in this way,
shown in Fig.~\ref{deteff}, is fit to a second-order polynomial.
This efficiency ($\varepsilon_{MC}$) must be corrected to
account for data-MC simulation differences in detector response:
\begin{equation}
\varepsilon=\varepsilon_{MC}/\Pi (1+\delta_i),
\label{eff}
\end{equation}
where $\delta_i$ are efficiency corrections, for each of several
effects. These corrections are discussed below and summarized in
Table~\ref{effcorr}. They are determined at the $\omega$, $\phi$,
and $J/\psi$ masses, where the relative level of background is small,
and a linear interpolation between their values is used for the
mass ranges between the $\omega$, $\phi$, and $J/\psi$.

Our preliminary selection contains a cut on the energy deposited by 
charged pions in the calorimeter ($E_{EMC}/p<0.9$), which is not simulated
accurately. The momentum dependence of the probability for a pion to 
have $E_{EMC}/p>0.9$ is found using events for the process 
$e^+e^-\to\pi^+\pi^-\pi^+\pi^-\gamma$ 
selected without cuts on energy deposition in the calorimeter.
The value of the efficiency correction is about 3\%.

The efficiency correction for the background        
suppression cuts is determined from ratios of the number of events selected 
with and without the background suppression cuts, in data and MC simulation.
We use events with $\chi^2<20$ from mass ranges near the $\omega$, $\phi$,
and $J/\psi$. In data the fraction of signal events
rejected by these cuts varies from 13\% in the $\omega$
and $\phi$ mass region to 20\% at $J/\psi$. This dependence is
reproduced by the simulation.  The efficiency correction is
about 3\% for all masses.

\begin{figure}
\includegraphics[width=0.47\textwidth]{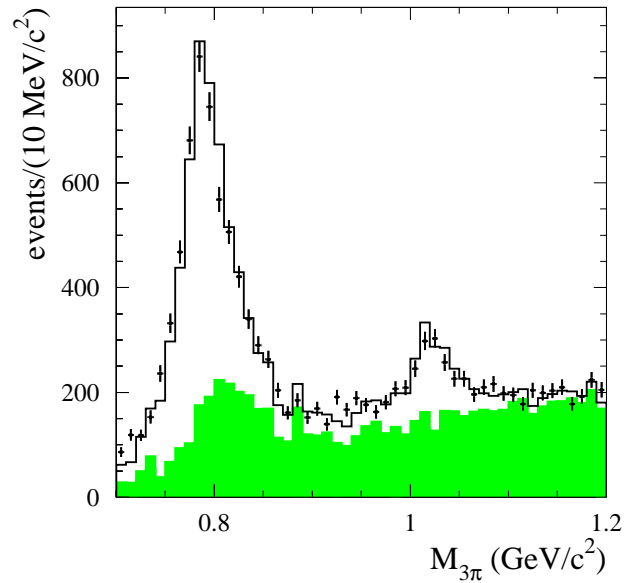}
\caption{The $3\pi$ mass spectrum for events with $40<\chi^2<500$. 
Points with error bars show the data distribution. 
The solid line shows the
sum of the distributions for simulated $e^+e^-\to\pi^+\pi^-\pi^0\gamma$
and background processes.
The shaded histogram shows the distribution for background processes.}
\label{bkg_test}
\end{figure}
To determine the efficiency correction due to the $\chi^2<40$ cut, we fit 
the $3\pi$ mass spectrum in the $\phi-\omega$ mass range for events
with $\chi^2 < 40$, $40 < \chi^2 < 500$, and $ 500 < \chi^2 < 1000$ to a sum
of simulated signal and background distributions with free scale factors.
The fit for events with $40<\chi^2<500$ is shown in Fig.~\ref{bkg_test}.
The data-MC simulation discrepancy is estimated by the double ratio
$$g(500)=\frac{N_{sig}(40 < \chi^2 < 500)/N_{MC}(40 < \chi^2 < 500)}{N_{sig}(\chi^2 < 40)/N_{MC}(\chi^2 < 40)},$$
where $N_{MC}$ is the number of simulated events and $N_{sig}$ is the number 
of signal events in data obtained from the fit to $M_{3\pi}$ in the 
corresponding $\chi^2$ interval.
We obtain $g(500)=1.30 \pm 0.04$ and, for the $500 < \chi^2 < 1000$ interval,
$g(1000)=1.29\pm 0.17$. For higher $\chi^2$ the relative signal 
level is too small to determine its value.
From the measured values of $g(500)$ and $g(1000)$ and simulated $\chi^2$
distributions, we calculate the efficiency correction for the $\omega$--$\phi$
mass region to be $(9\pm3)\%$.
The quoted error includes uncertainties due to errors on the $g$ values
and the uncertainty due to events with $\chi^2>1000$. For these events
the $g$ value found for the $500<\chi^2<1000$ interval is used.
In the $J/\psi$ mass region we use a mass-sideband subtraction
method to determine the numbers of signal events with $\chi^2<40$ and
$40<\chi^2<500$ in data and MC simulation. For higher $\chi^2$ we do not 
see a $J/\psi$ signal due to large background levels. 
The obtained value of $g(500)=1.12\pm0.20$ agrees with the result 
for the $\phi$-$\omega$ mass region. Using this number for all 
$\chi^2$ we find the efficiency correction at the $J/\psi$ mass to be 
$(4\pm 6)\%$.

\begin{figure}
\includegraphics[width=0.47\linewidth]{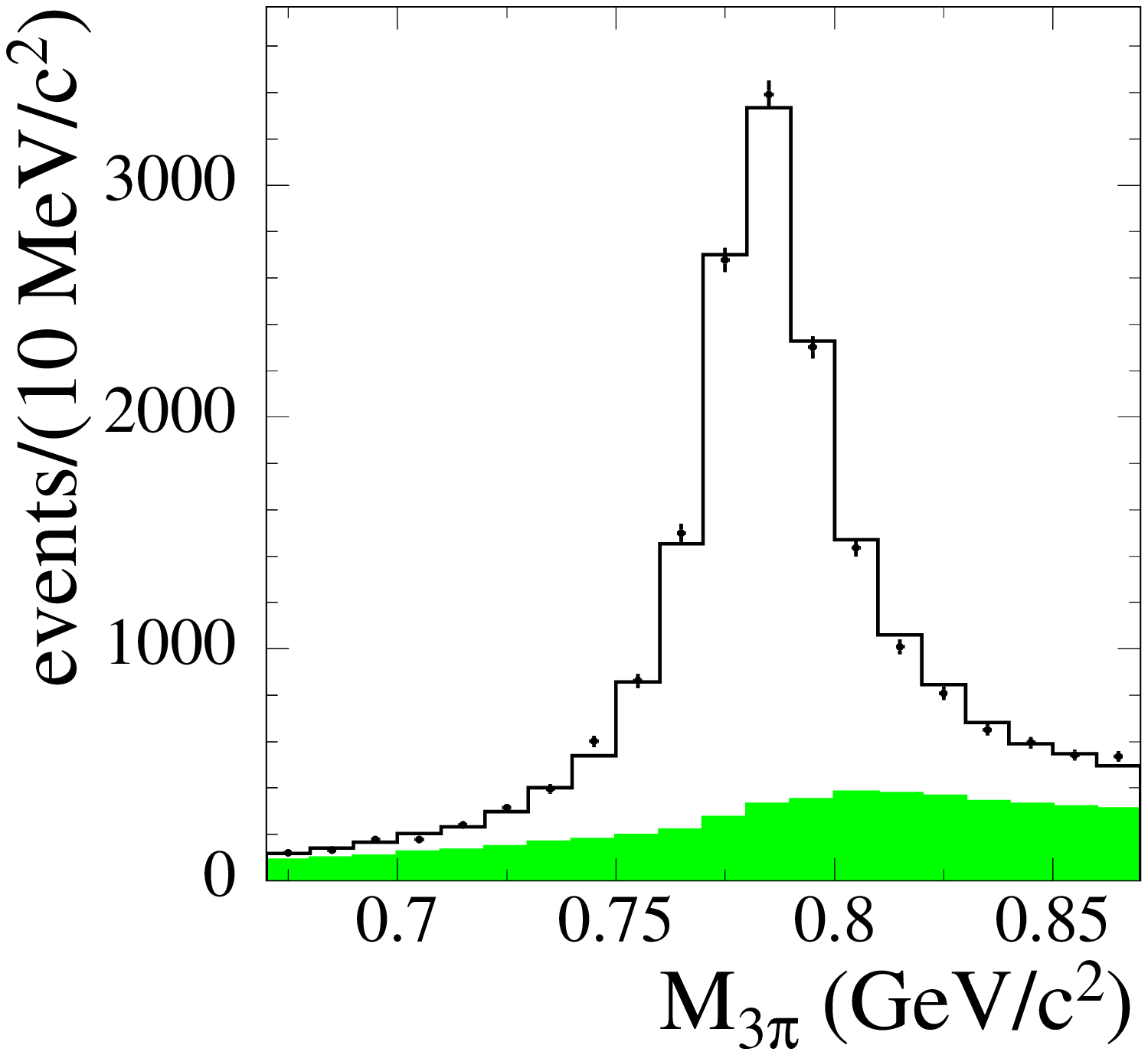}
\includegraphics[width=0.47\linewidth]{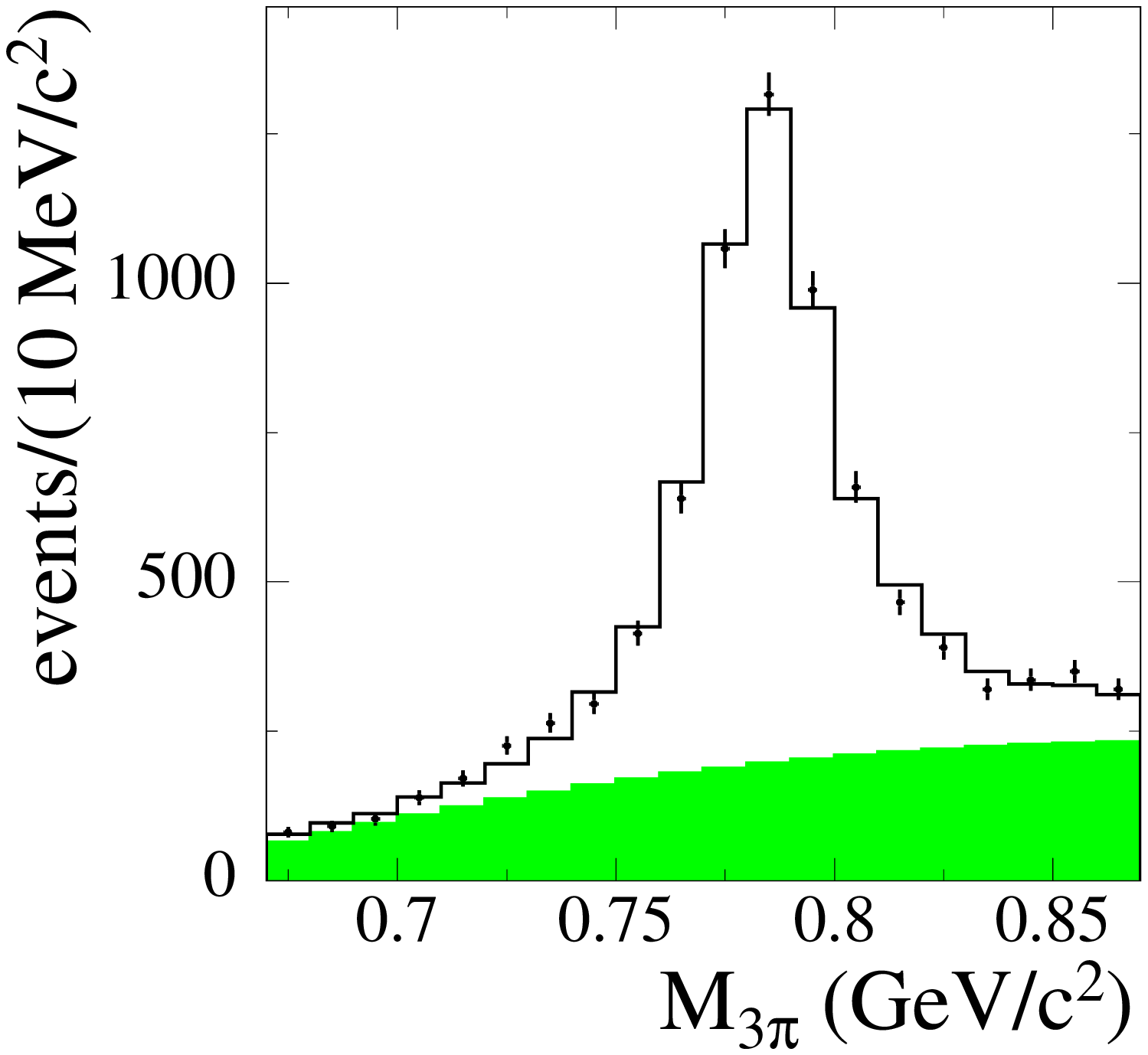}
\caption{The $3\pi$ mass distribution for events
selected without using the photons from the $\pi^0$ decay,
that are reconstructed (left) and not reconstructed (right) with 
our standard kinematic fit procedure.
Points with error bars show the data distribution. The solid line is
a fit result. The shaded histogram is the fitted background contribution.}
\label{pi0loss}
\end{figure}

The other possible source of data-MC simulation difference is $\pi^0$ losses
due to the merging of electromagnetic showers of the two photons from the
$\pi^0$ decay or the loss of one of the decay photons.
To study the $\pi^0$ losses we perform
a kinematic fit for data and simulated events with the
$e^+e^- \to \pi^+\pi^-\pi^0\gamma$ hypothesis using the measured parameters
for only the two charged tracks and the ISR photon.  The $\pi^0$ energy and
angles are determined as a result of the fit. To suppress background we use
a very tight cut on the $\chi^2$ of the fit
and require that the total energy of all photons in the event, excluding
the ISR photon candidate,
is greater than 80\% of the $\pi^0$ energy found in the fit. The high level of
remaining background does not allow determination of the efficiency correction for
masses above the $\omega$. Therefore, we restrict our study to the
$\omega$-mass region. The $3\pi$ mass spectra
for selected events reconstructed ($\chi^2 < 10000$) 
and not reconstructed with our standard
kinematic fit procedure are shown in Fig.~\ref{pi0loss}.
The fraction of events that are not reconstructed is about 30\%.
In most of these events one of the photons from the $\pi^0$
decay is outside the polar angle range used in our standard selection.
The mass spectra are fit to a sum of distributions
for signal and background events. The signal distribution is extracted from
the simulation. The background spectrum is a sum of the simulated distribution
for $e^+e^- \to \pi^+\pi^-\pi^0\pi^0\gamma$ events and a second order
polynomial with free coefficients. The efficiency correction due to $\pi^0$
losses is
determined to be $\delta=\epsilon_{MC}/\epsilon_{exp}-1=-(1.9\pm0.9)\%$.
Here $\epsilon$ is the fraction of events with a $\pi^0$
reconstructed with the constrained fit discussed above that,
after the standard kinematic fit procedure, pass the $\chi^2 < 10000$
selection.
To calculate the correction for higher
masses we must take into account the dependence of the shape of the $\pi^0$ 
energy spectrum on the $3\pi$ mass. At the $\omega$ mass
we determine the correction as a function of $\pi^0$ energy and convolve
this function with the $\pi^0$ energy spectra at the $\phi$ and
$J/\psi$ masses. The calculated corrections are
$-(1.7\pm0.9)\%$ and $-(1.5\pm0.8)\%$ for $\phi$ and $J/\psi$ masses,
respectively.

We also studied the quality of the simulation of trigger and background
filters used in event reconstruction. The corresponding 
efficiency corrections are listed in Table~\ref{effcorr}.
We use the overlap of the samples of events passing either different filters
or trigger criteria and the partial independence of these filters or
triggers to measure their efficiency.

The data-MC simulation difference in track losses is studied by comparing 
the ratios of $e^+e^-\to\pi^+\pi^-\pi^+\pi^-\gamma$ events
with three and four tracks in data and MC simulation. The difference in data 
and simulated probabilities to lose one of four tracks is found to be 
$(3.6\pm2.0)\%$.
For the case of two tracks we estimate the corresponding efficiency
correction to be $(1.8\pm1.8)\%$. We increase the systematic error to
account for the possible dependence of the correction on track multiplicity
and track momenta.

The data-MC simulation difference in the probability of photon conversion in 
the detector material before the DCH is studied using
$e^+e^-\to \gamma\gamma$ events and is found
to be $-(0.4\pm0.2)\%$. We estimate that the total correction for conversion of
one of the three photons in an event is $-(1.0\pm0.6)\%$.
The fact that part of this correction is already included in the
correction for $\pi^0$ loss is accounted for in the determination of this
value.

\begin{figure}
\includegraphics[width=0.48\linewidth]{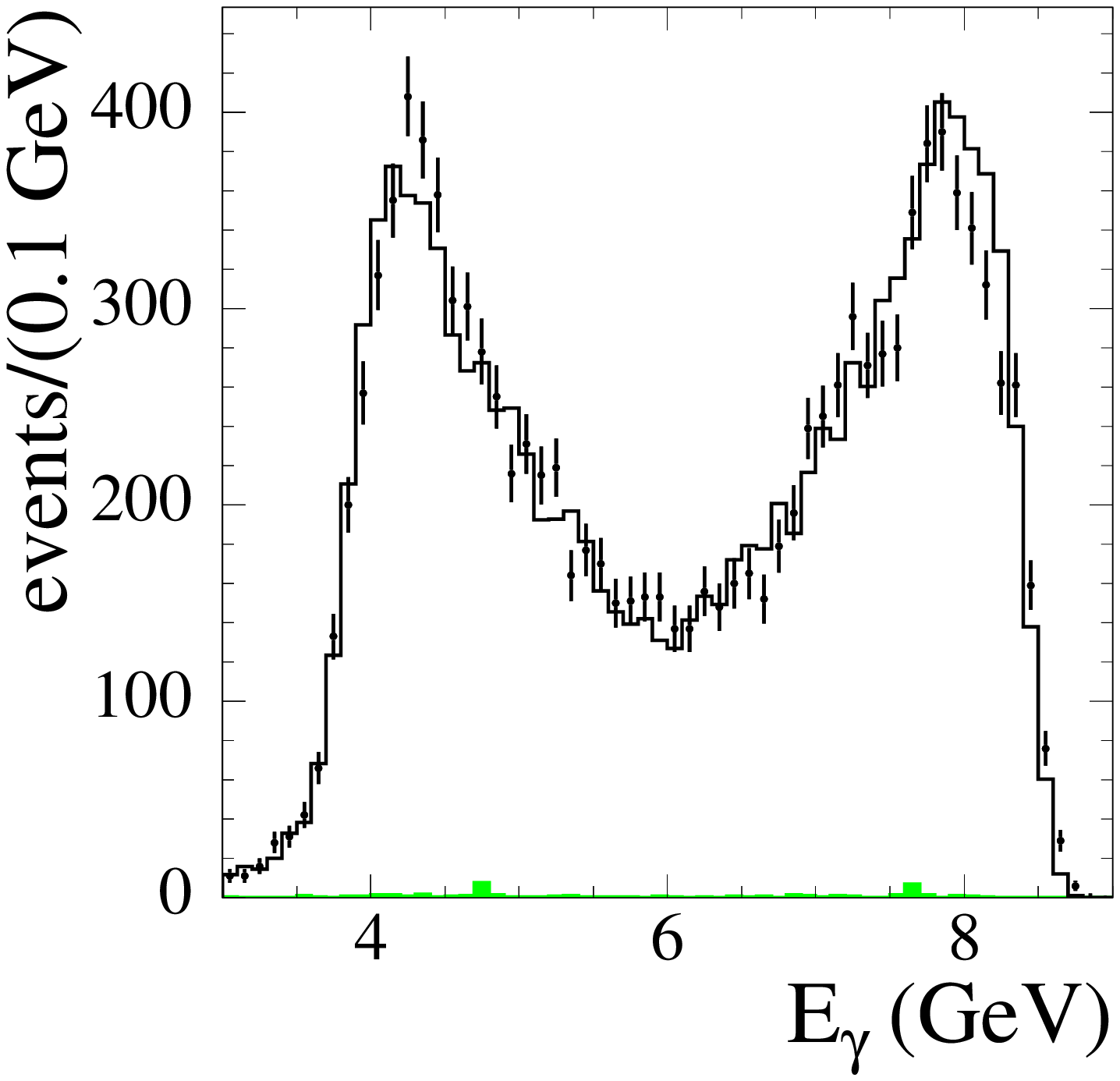}
\hfill
\includegraphics[width=0.48\linewidth]{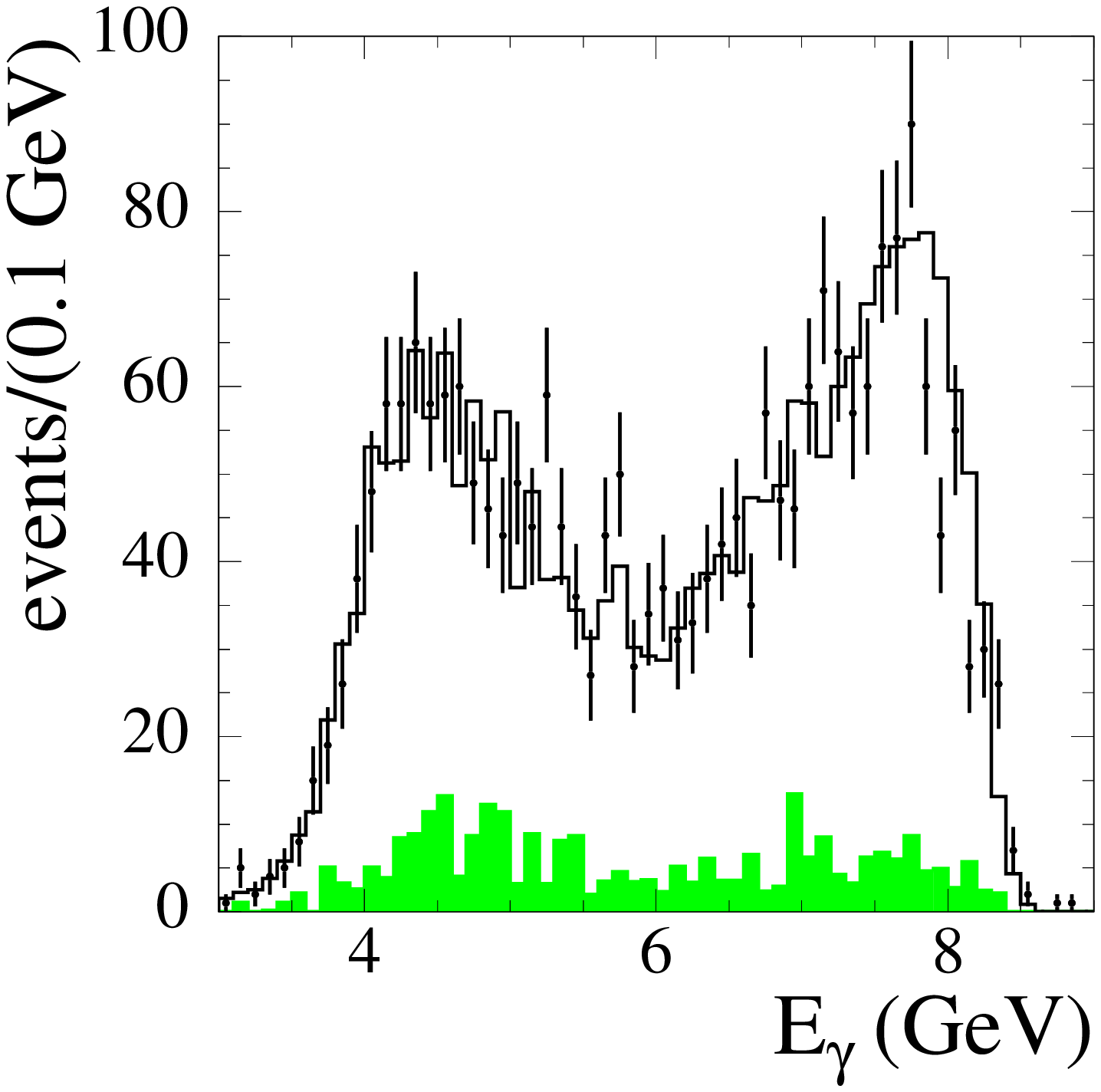}
\vspace{1mm}\\
\includegraphics[width=0.48\linewidth]{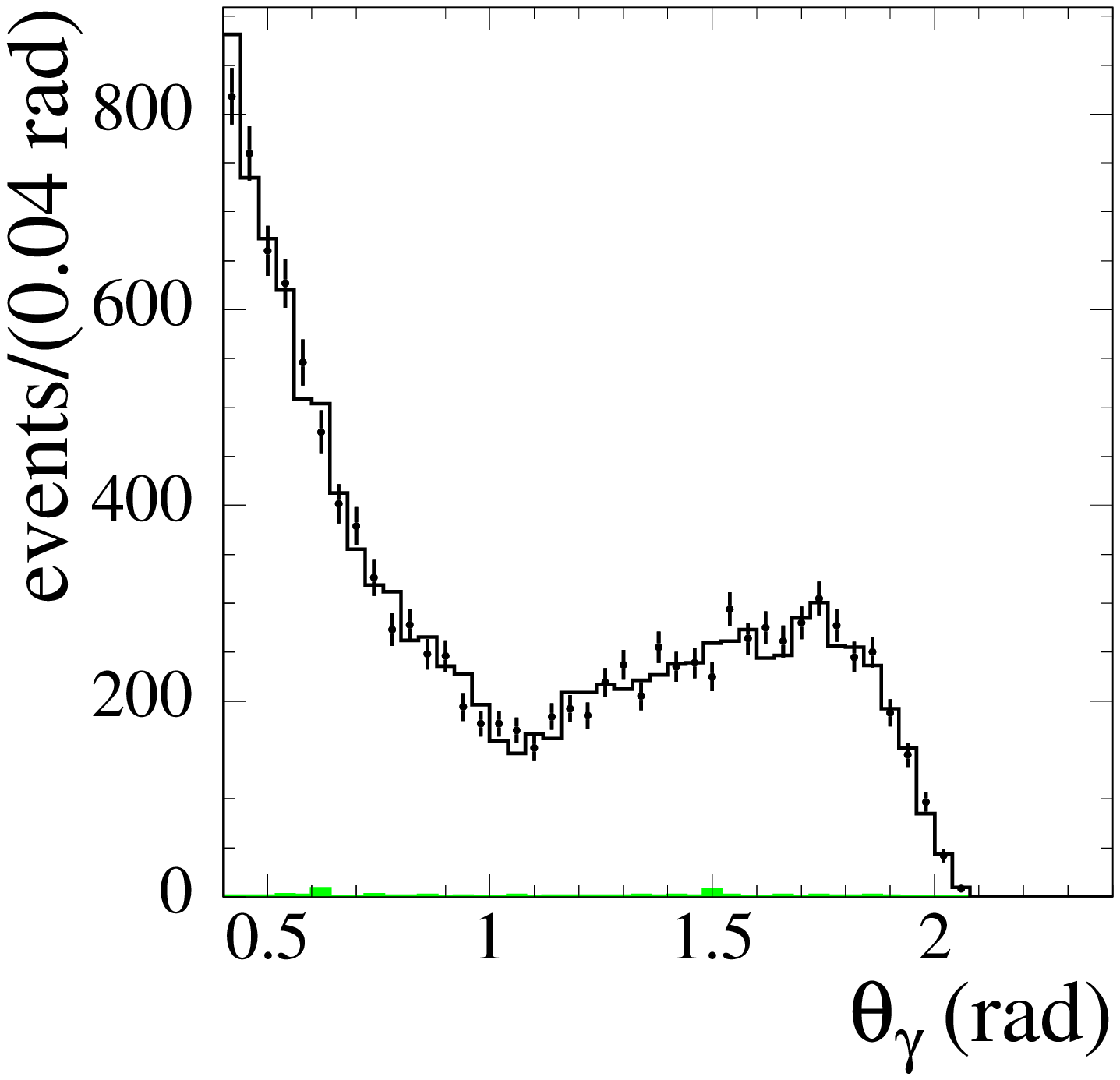}
\hfill
\includegraphics[width=0.48\linewidth]{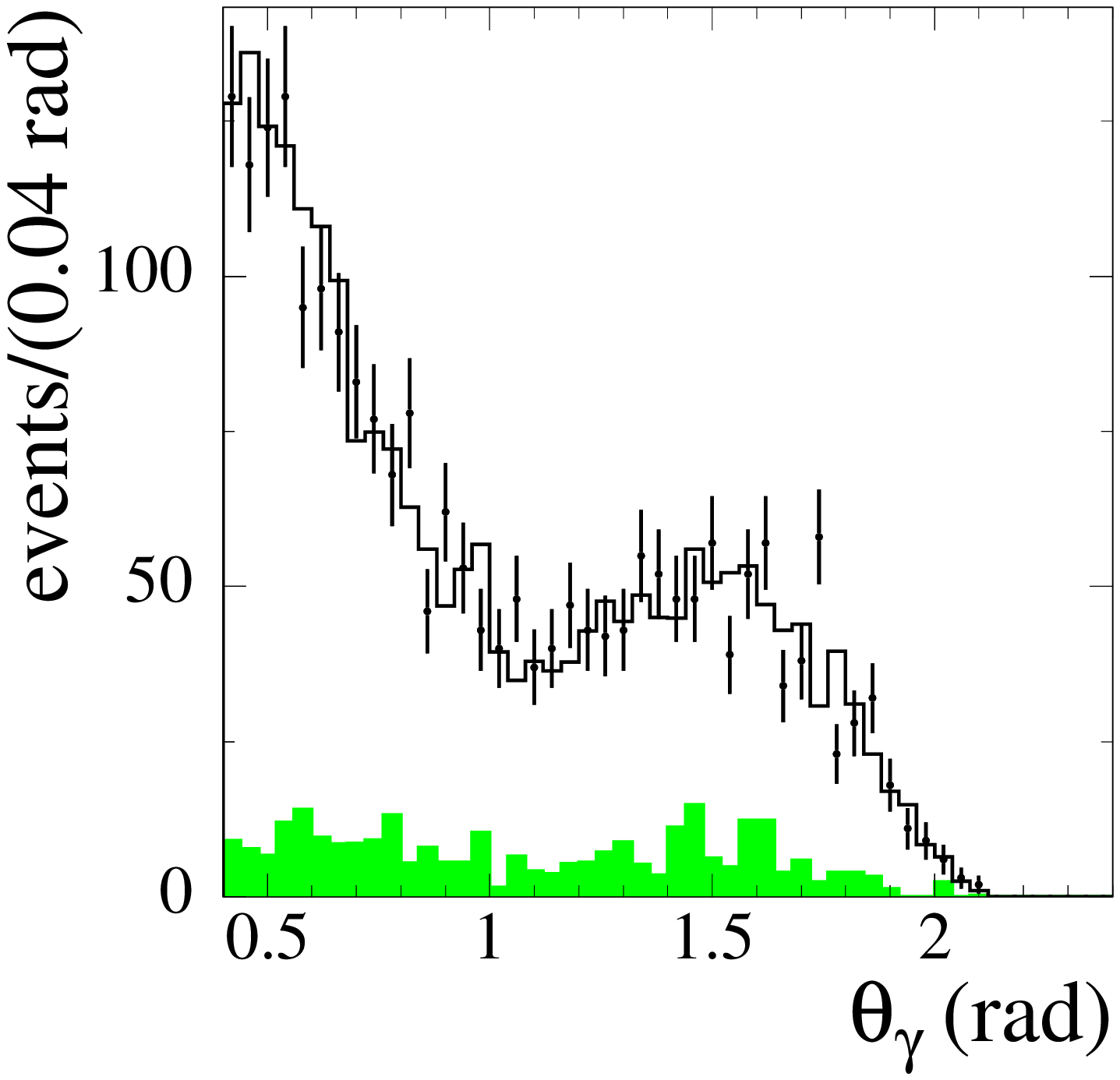}
\caption{Distributions of the ISR photon energy (1st row) and
polar angle (2nd row) for data (points with error bars)
and simulation (solid line) events with $\chi^2<20$. 
The left and right columns correspond to the mass
regions $0.75<M_{3\pi}<0.82$ GeV/$c^2$ and
$1.40<M_{3\pi}<1.80$ GeV/$c^2$, respectively. Shaded histograms show
the calculated background contributions.} 
\label{difdis0}
\end{figure}
\begin{figure}
\includegraphics[width=0.48\linewidth]{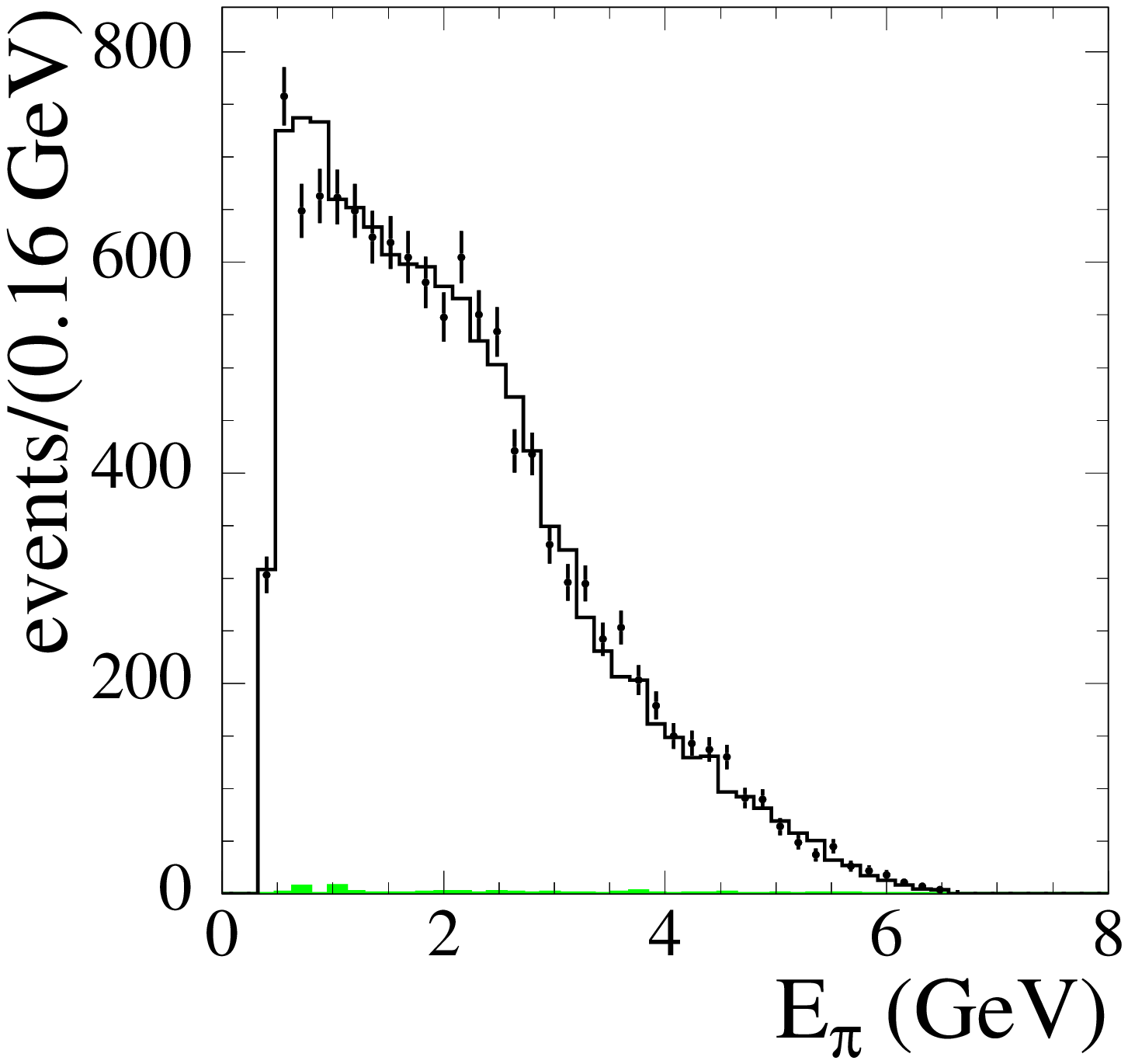}
\hfill
\includegraphics[width=0.48\linewidth]{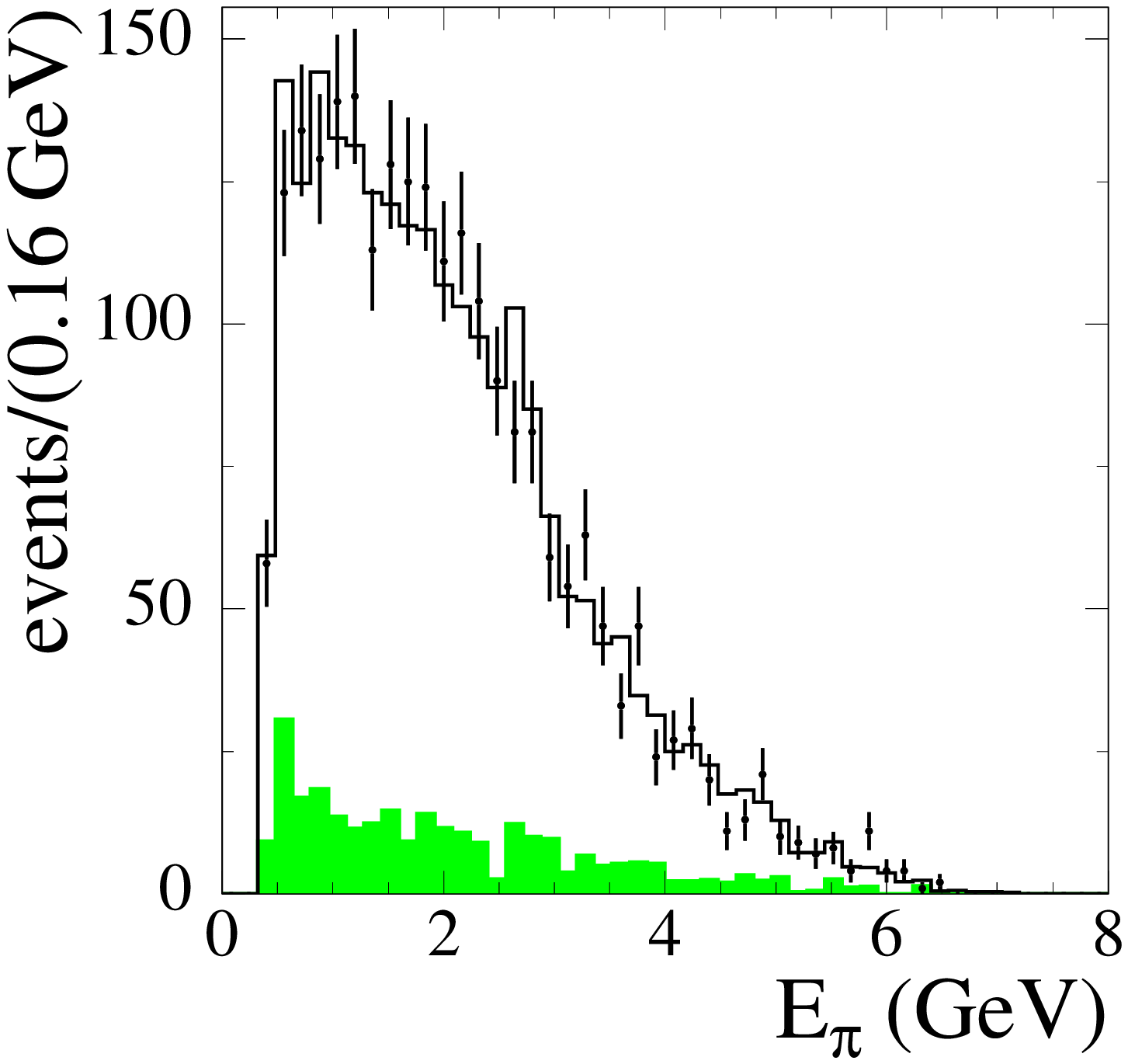}
\vspace{1mm}\\ 
\includegraphics[width=0.48\linewidth]{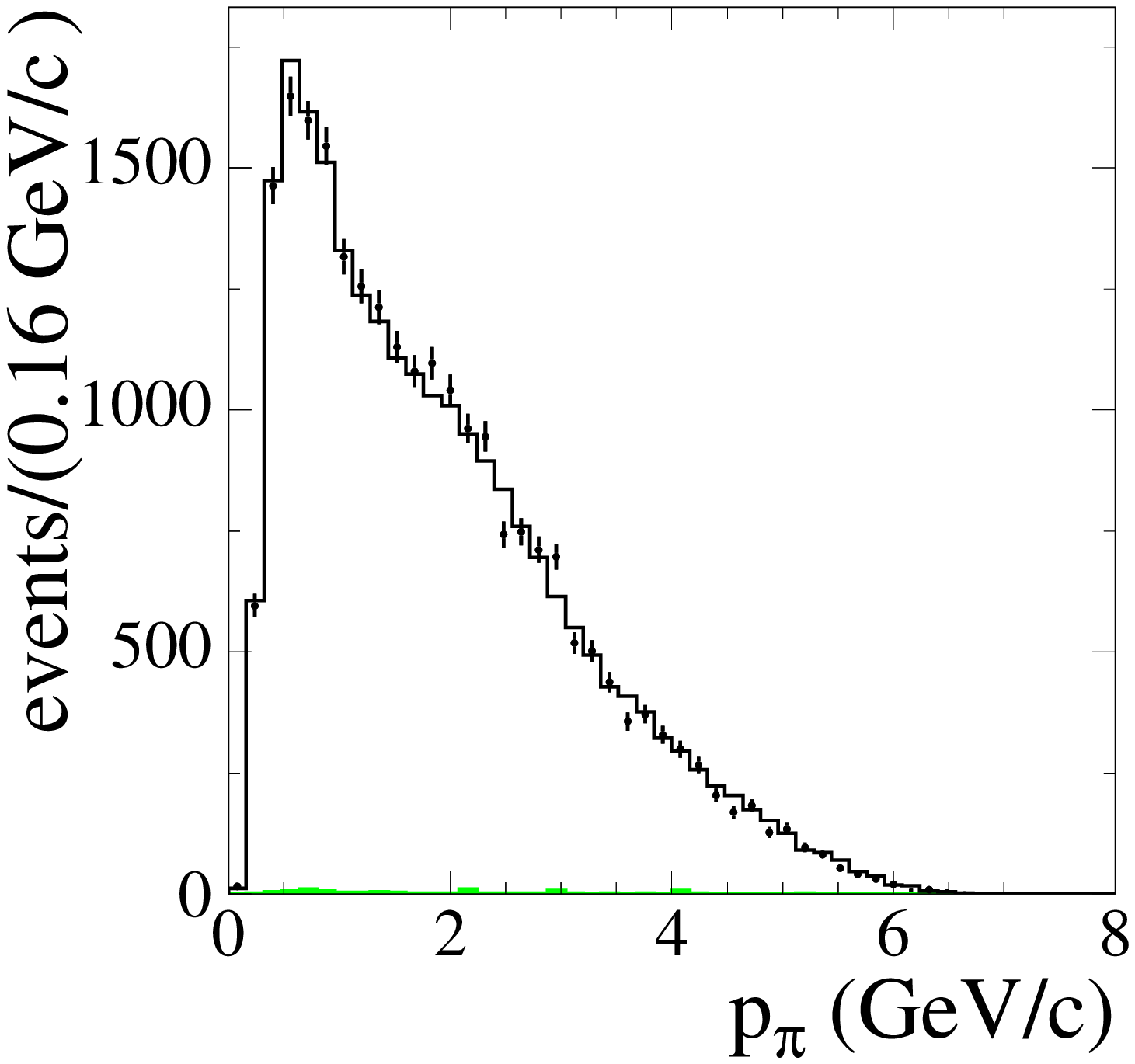}
\hfill
\includegraphics[width=0.48\linewidth]{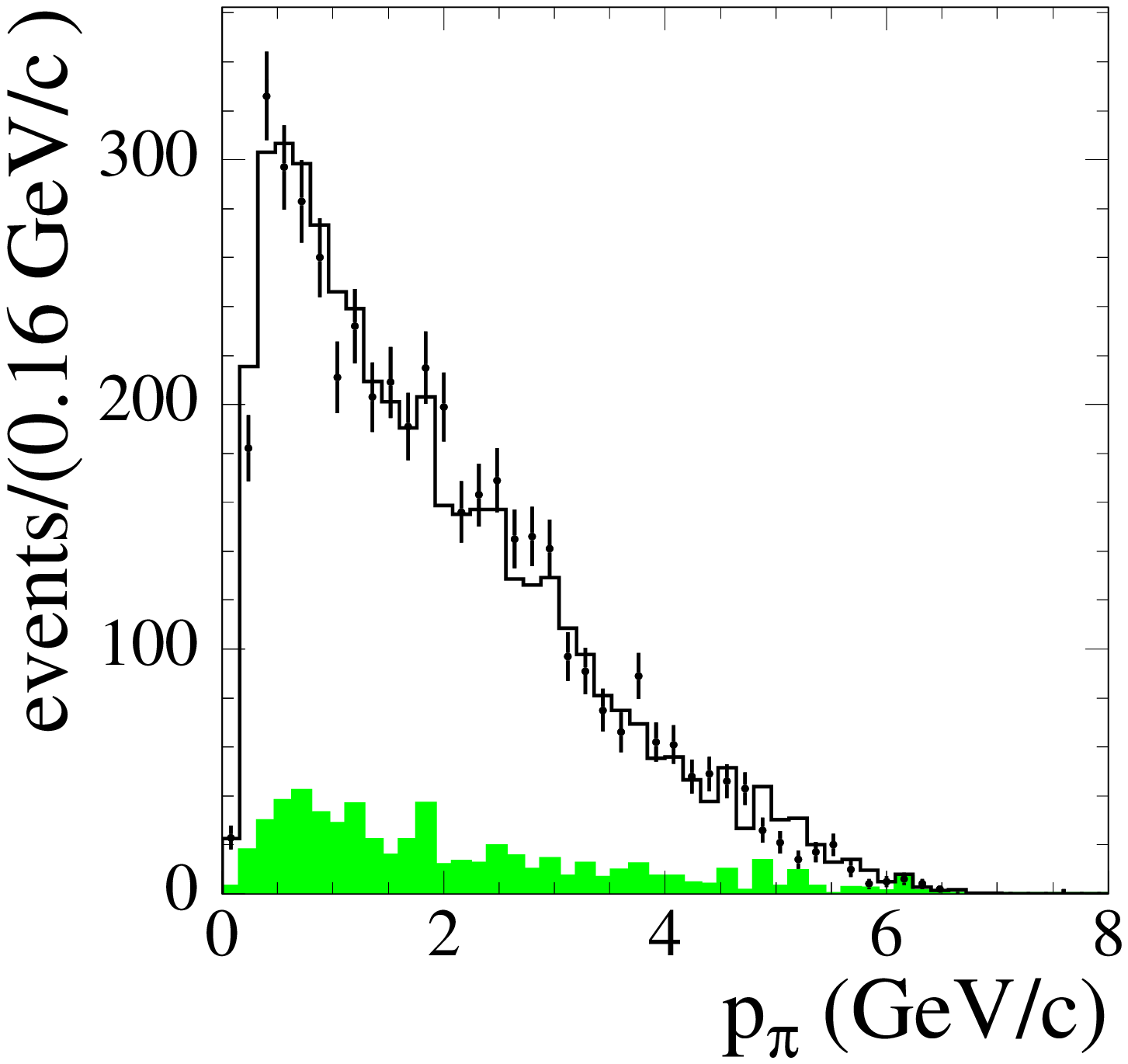}
\vspace{1mm}\\
\includegraphics[width=0.48\linewidth]{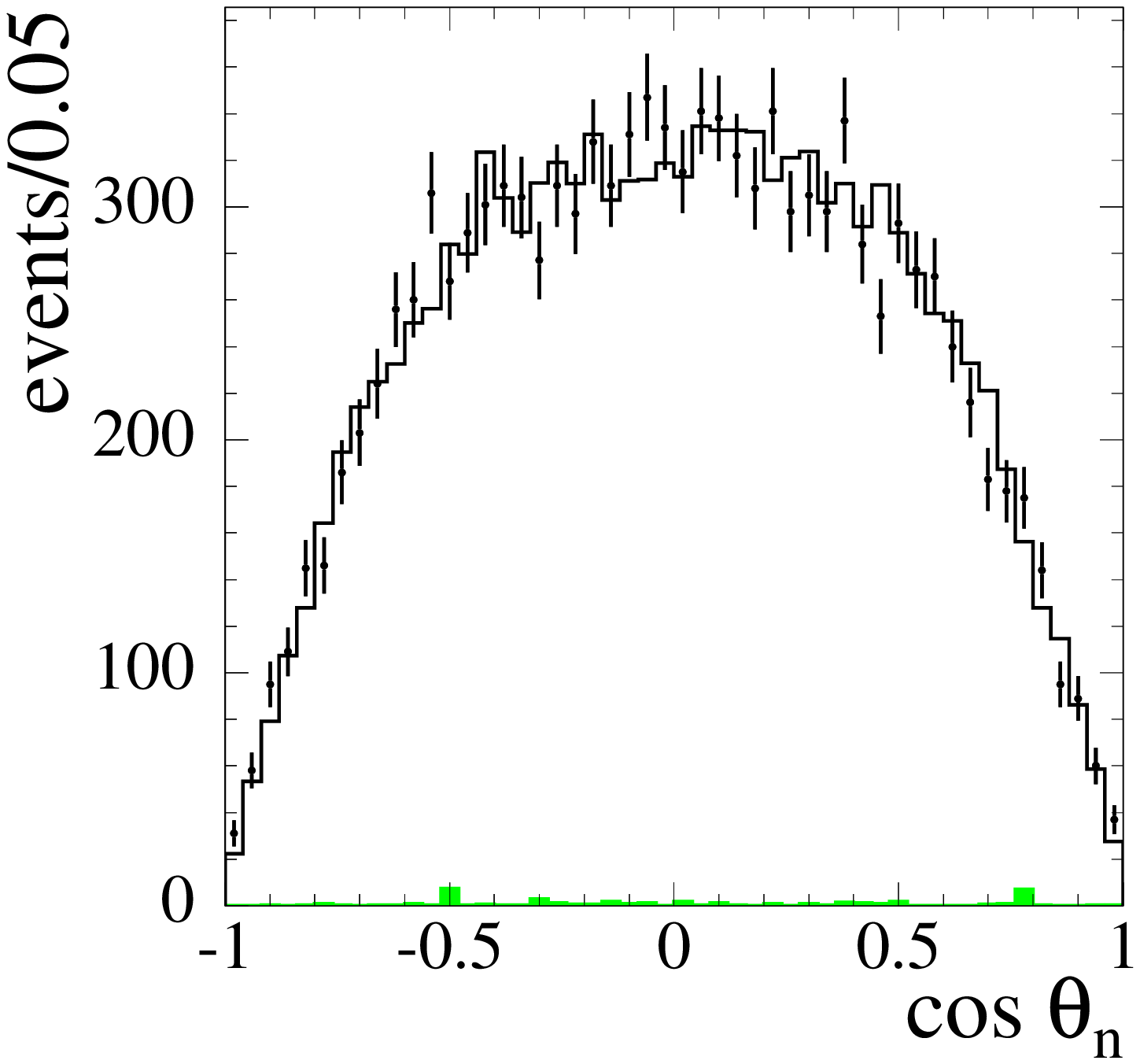}
\hfill
\includegraphics[width=0.48\linewidth]{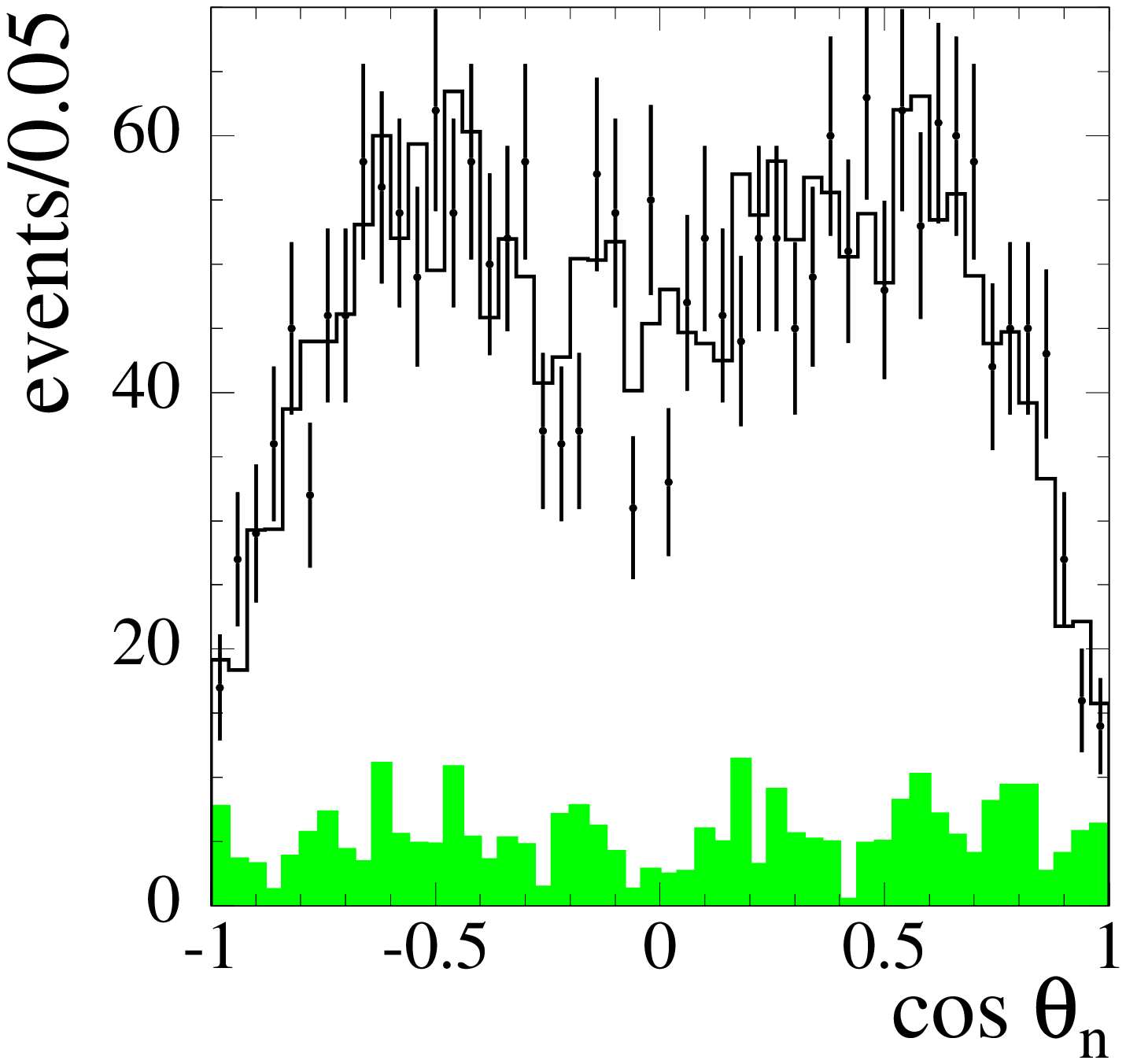}
\caption{Distributions of 
$\pi^0$ energy (1st row), charged pion momentum (2nd row),
and angle between the normal to the $3\pi$ plane
in the three-pion rest frame and the direction of the $3\pi$
system in the laboratory frame
(3rd row) for data (points with error bars)
and simulation (solid line) events with $\chi^2<20$. The left and right
columns correspond to the mass
regions $0.75<M_{3\pi}<0.82$ GeV/$c^2$ and
$1.40<M_{3\pi}<1.80$ GeV/$c^2$, respectively. Shaded histograms show
the calculated background contributions.} 
\label{difdis1}
\end{figure}
\begin{figure}
\includegraphics[width=0.48\linewidth]{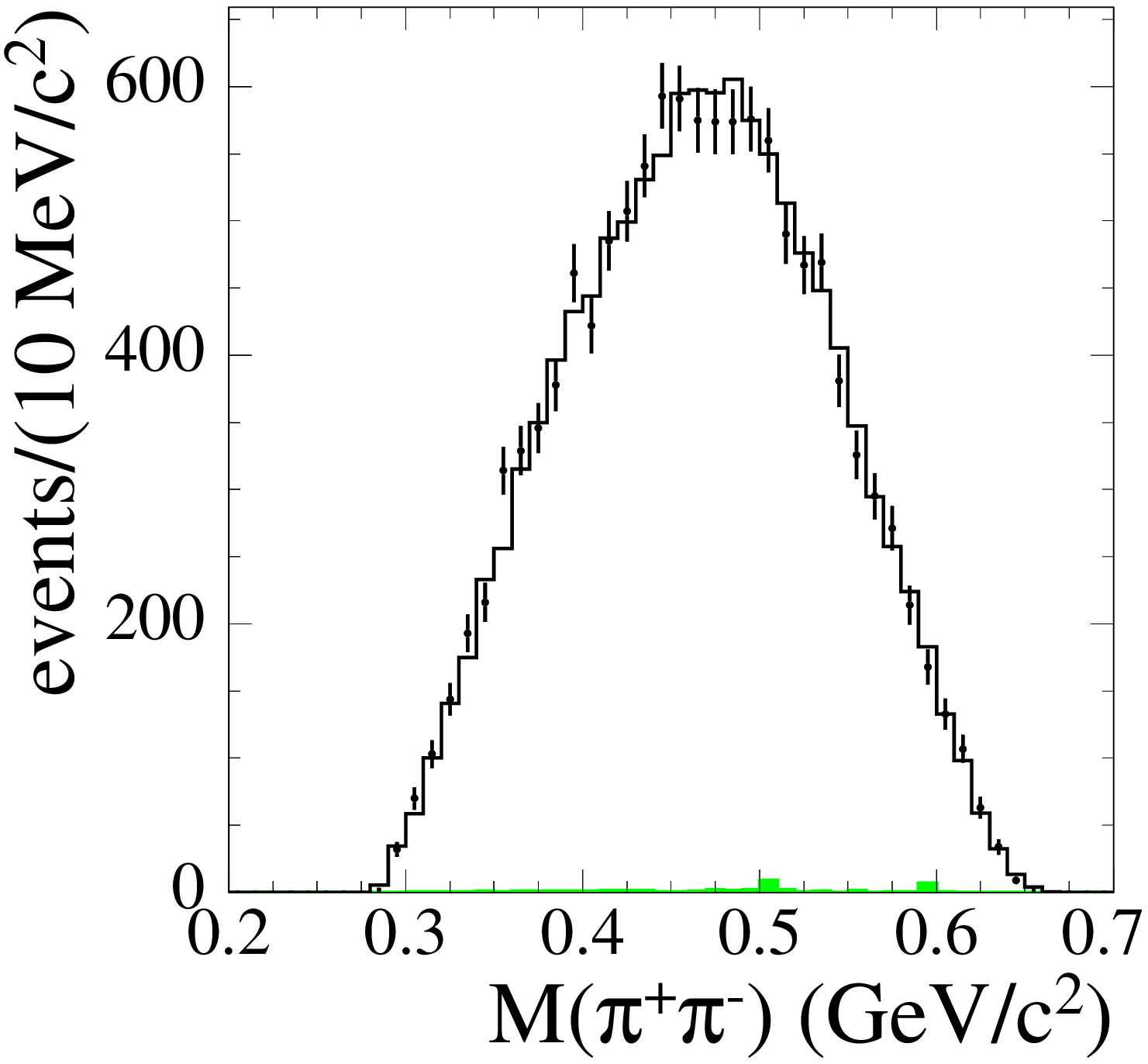}
\hfill
\includegraphics[width=0.48\linewidth]{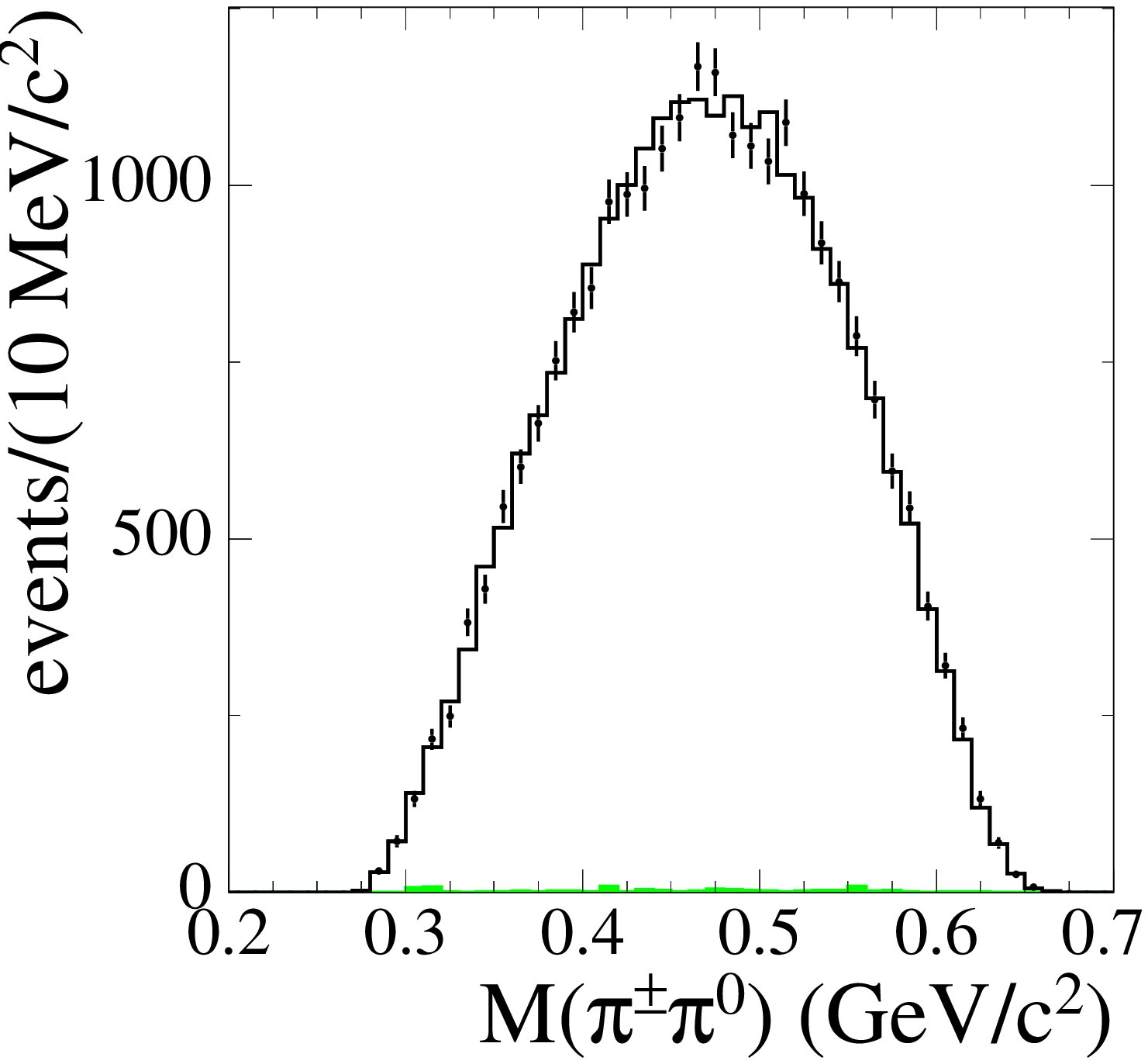}
\vspace{1mm}\\
\includegraphics[width=0.48\linewidth]{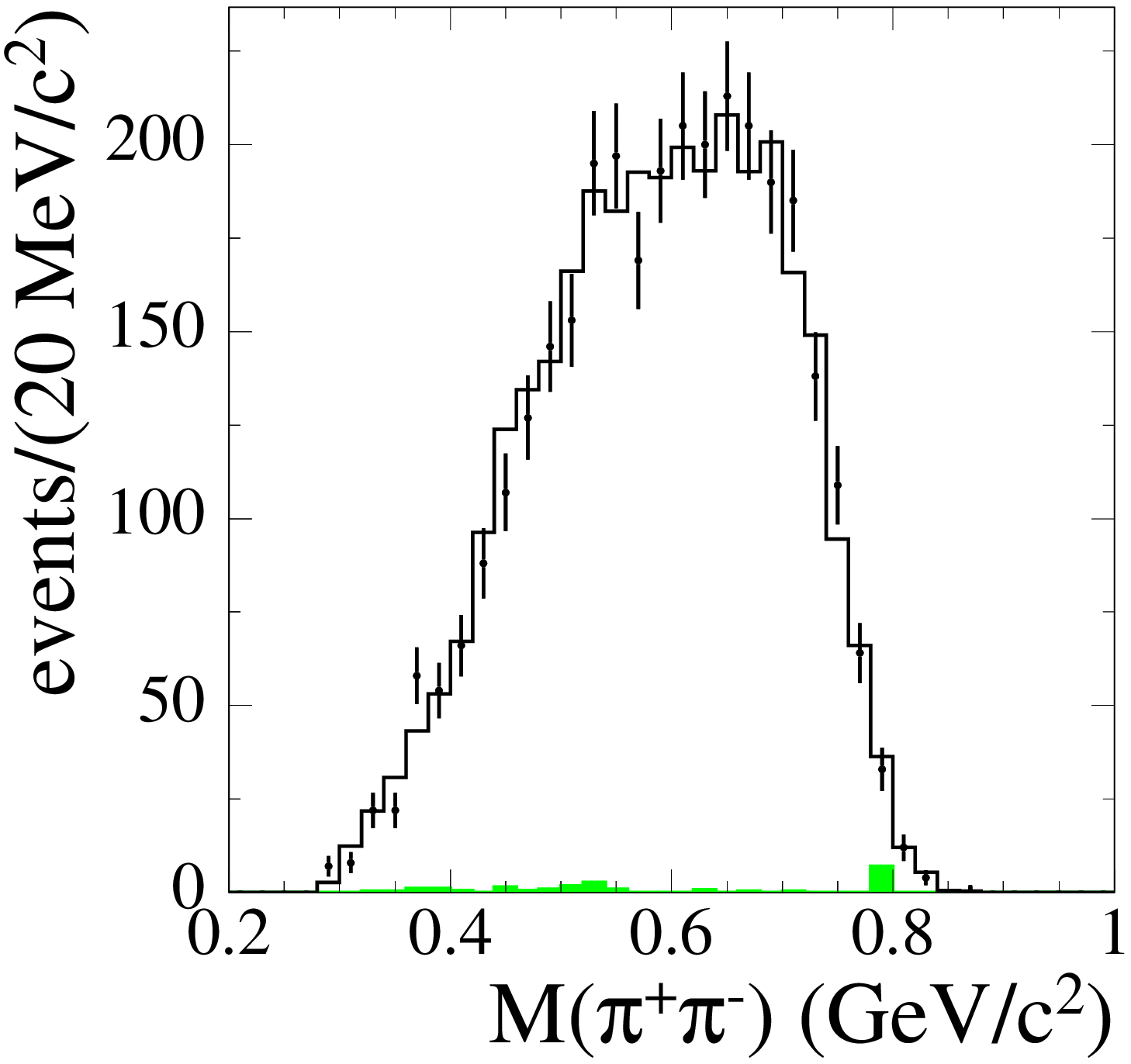}
\hfill
\includegraphics[width=0.48\linewidth]{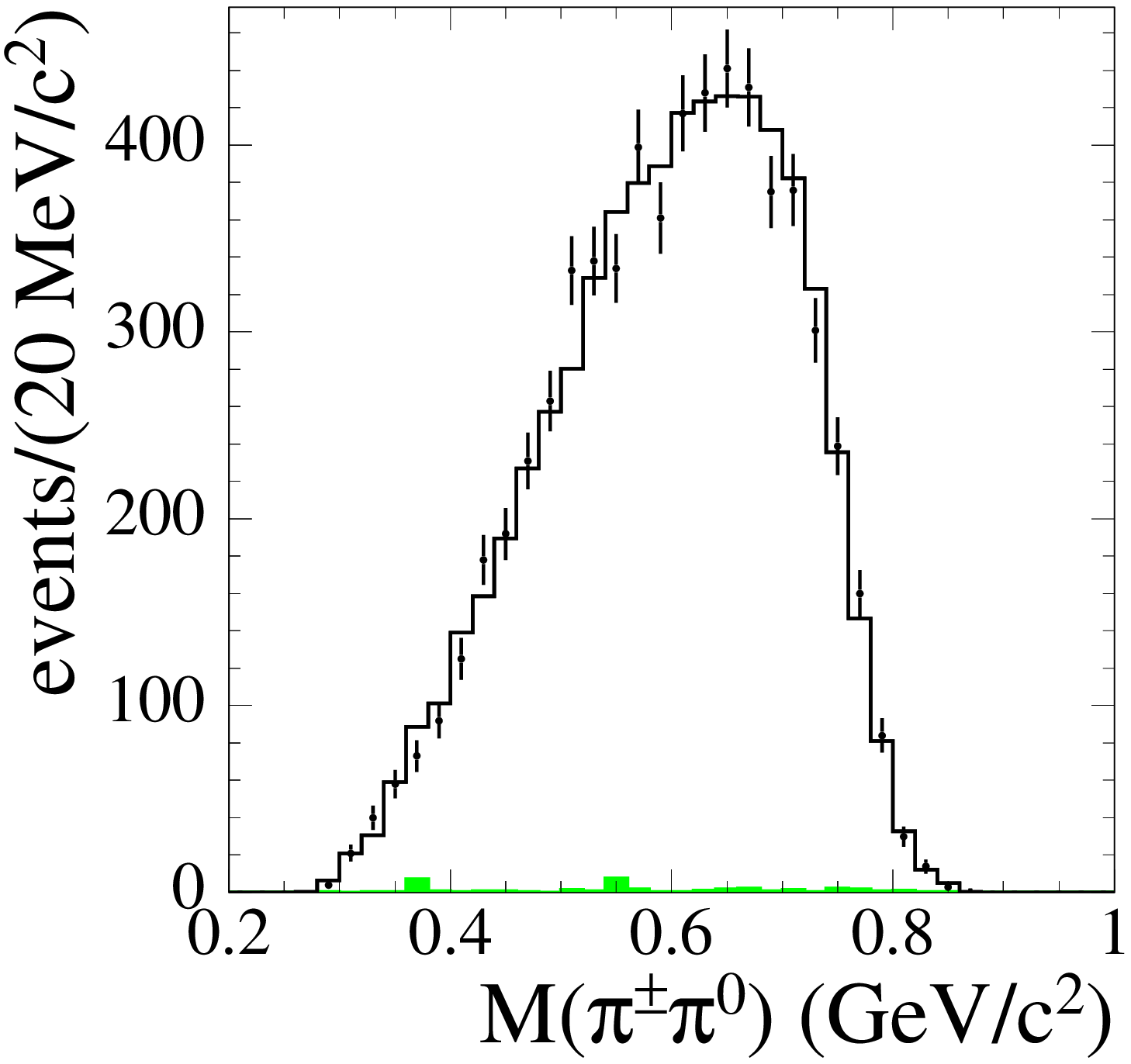}
\vspace{1mm}\\
\includegraphics[width=0.48\linewidth]{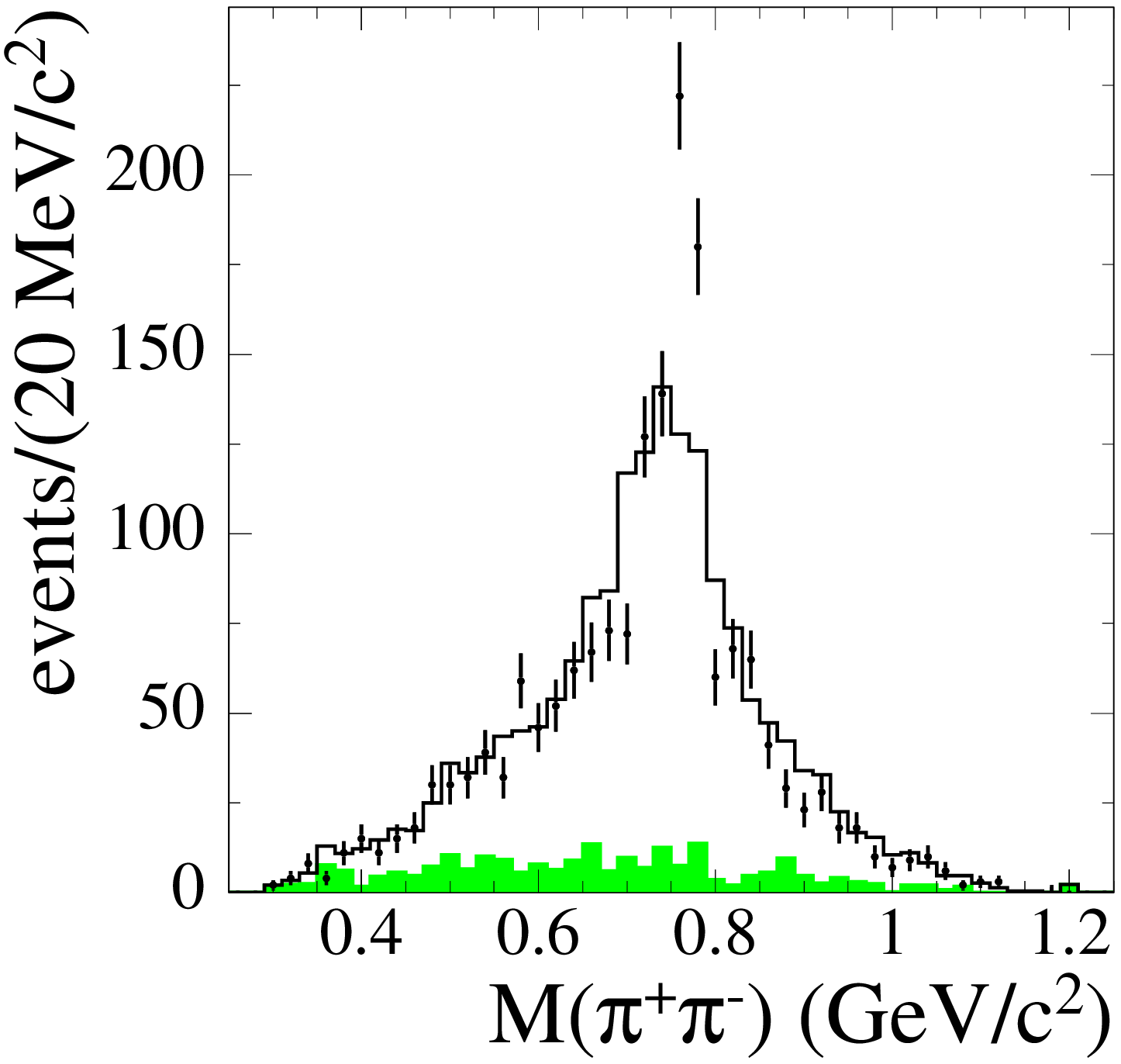}
\hfill
\includegraphics[width=0.48\linewidth]{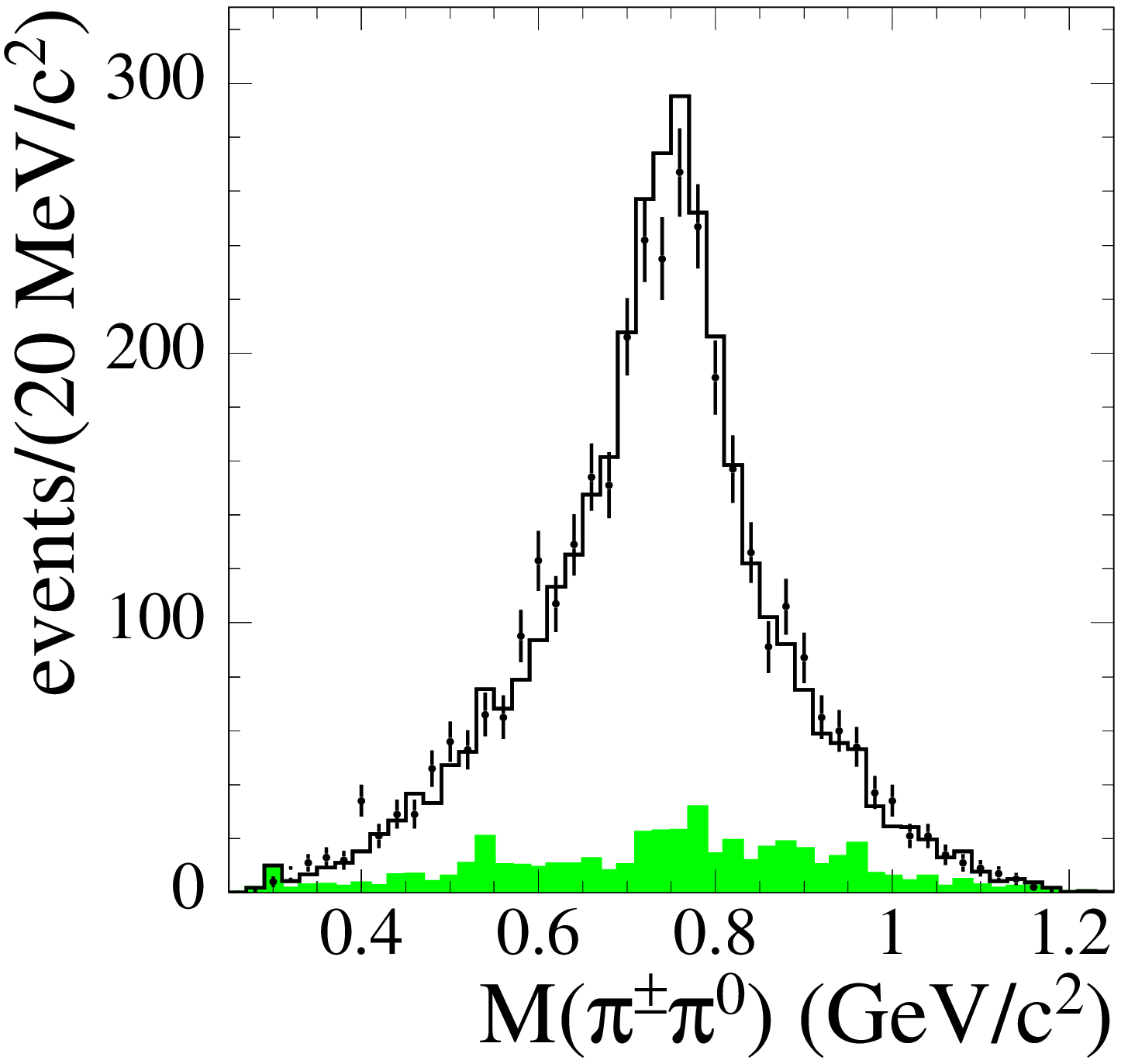}
\vspace{1mm}\\
\includegraphics[width=0.48\linewidth]{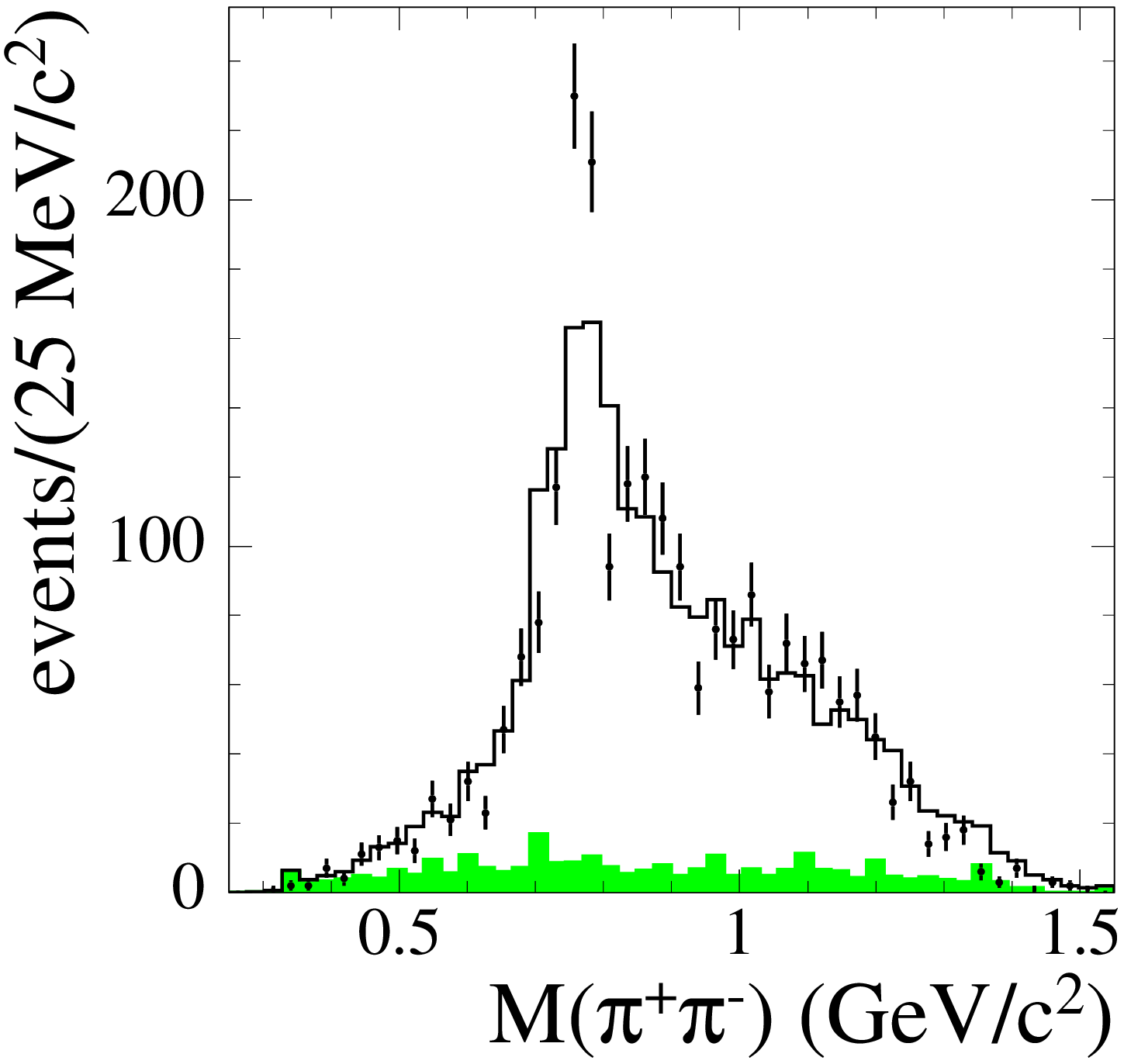}
\hfill
\includegraphics[width=0.48\linewidth]{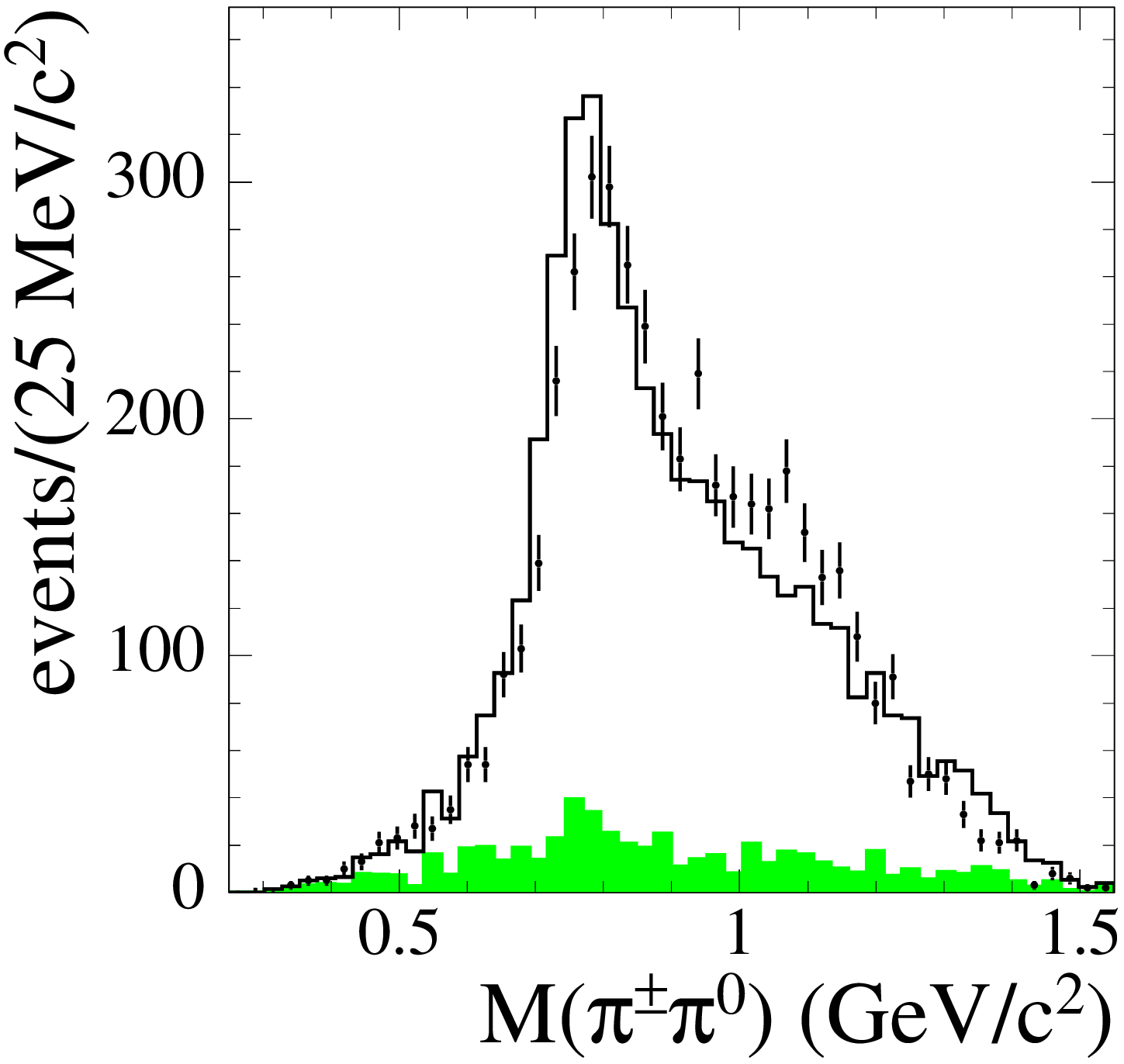}
\caption{Distributions of $\pi^+\pi^-$ (left column) and $\pi^\pm\pi^0$
(right column) invariant masses for data (points with error bars) and
simulation (solid line) for events with $\chi^2<20$. The rows correspond to
the mass regions: $0.75<M_{3\pi}<0.82$ GeV/$c^2$ (1st row),
$1.00<M_{3\pi}<1.04$ GeV/$c^2$ (2nd row), $1.10<M_{3\pi}<1.40$ GeV/$c^2$
(3rd row), and $1.40<M_{3\pi}<1.80$ GeV/$c^2$ (4th row).
Shaded histograms show the calculated background contributions.} 
\label{difdis2}
\end{figure}

The event generator for the $e^+e^-\to 3\pi\gamma$ reaction uses a
model with an intermediate $\rho\pi$ state. This model has been
checked for the $\omega-\phi$ mass region in several high statistics $e^+e^-$
experiments~\cite{SND2003,SND2001,DAFNE}. Small deviations
from the $\rho\pi$ model found in the $\phi$ meson decays~\cite{DAFNE} are
negligible at our statistical level. Figures~\ref{difdis0}, \ref{difdis1},
and \ref{difdis2}
demonstrate a good agreement between the data and simulated distributions 
over both the kinematical and dynamical (di-pion invariant masses) parameters
for the $\omega$ mass region.

In the mass region above the $\phi$ the intermediate state $\omega\pi$
becomes noticeable. This additional mechanism was studied at the SND 
experiment~\cite{SND2002}
for the energy region up to 1.4 GeV. It was established that the contribution
of the $\omega\pi$ intermediate state to the total cross section of
$e^+e^-\to 3\pi$ process does not exceed 10\%. 

The distributions
of different kinematic parameters for the mass region
from 1.4 to 1.8 GeV/$c^2$
are shown in Figs.~\ref{difdis0} and \ref{difdis1}.
For these parameters the data distributions agree with the simulated ones.
In the $\pi^+\pi^-$ mass spectra (Fig.~\ref{difdis2}) for data, a narrow
peak near the $\omega$ mass is seen. The fraction of events in this peak is
about 10\% for $1.1<M_{3\pi}<1.4$ GeV/$c^2$ and 6\% for
$1.4<M_{3\pi}<1.8$ GeV/$c^2$ in agreement with
measurements of SND~\cite{SND2002}. We calculate the detection efficiency
for simulated events with $0.77<M_{\pi^+\pi^-}<0.80$ GeV/$c^2$ and find
its difference with detection efficiency for the full $M_{\pi^+\pi^-}$ 
mass range to be $(1.6\pm3.9)\%$.

There are also noticeable
deviations from the MC simulation in the distribution of the $\pi^\pm\pi^0$
invariant mass for the mass region $1.4<M_{3\pi}<1.8$ GeV/$c^2$
(Fig.~\ref{difdis2}), which
are manifest in the shift of the visible $\rho$-meson mass position and a bump
at a mass above 1 GeV/$c^2$. A possible explanation of
this difference is the appearance at higher masses of
the $\omega^{\prime\prime} \to \rho^\prime\pi$ transition, which interferes
with the $\rho\pi$ amplitude. We study the dependence of the detection
efficiency on $\pi^+\pi^0$ mass and find it to be constant in the mass
range $0.65<M_{\pi^+\pi^0}<1.25$ GeV/$c^2$.
We conclude that the use of the $\rho\pi$ model for the simulation
of $e^+e^-\to 3\pi\gamma$
does not lead to any significant errors in the determination of the detection
efficiency.

\begin{table*}
\caption{The values of different efficiency corrections $\delta_i$
for $\omega$, $\phi$, and $J/\psi$ mass regions.}
\begin{ruledtabular}
\begin{tabular}{lccc}
effect & $\delta_i(m_\omega)$,\% & $\delta_i(m_\phi)$,\% & $\delta_i(m_{J/\psi})$,\% \\
\hline
$E_{EMC}/p<0.9$ cuts      & $+2.9\pm 0.3$ & $+2.9\pm 0.3$ & $+2.5\pm 0.3$ \\
background rejection cuts & $+3.0\pm 0.4$ & $+2.4\pm 0.7$ & $+2.9\pm 1.9$ \\
$\chi^2<40$ cut           & $+9\pm 3$     & $+9\pm 3$     & $+4\pm 6$     \\
$\pi^0$ loss              & $-1.9\pm 0.9$ & $-1.7\pm 0.9$ & $-1.5\pm 0.8$\\
trigger and filters       & $+0.0\pm 0.4$ & $+0.1\pm 1.0$ & $+2.2\pm 2.0$ \\
track loss                & $+1.8\pm 1.8$ & $+1.8\pm 1.8$ & $+1.8\pm 1.8$ \\
photon conversion         & $-1.0\pm 0.6$ & $-1.0\pm 0.6$ & $-1.0\pm 0.6$ \\
\hline
total                     & $+14\pm4$     & $+14\pm4$     &$+11\pm7$      \\
\end{tabular}
\end{ruledtabular}
\label{effcorr}
\end{table*}
Efficiency corrections $\delta_i$ are summarized in Table~\ref{effcorr}.
The total efficiency corrections $\delta$ calculated from 
$1+\delta=\Pi (1+\delta_i)$
are listed in the last row of the table.
The correction is 14\% near the $\omega$ and $\phi$, and
11\% at the $J/\psi$ mass.

\section{ \boldmath Fit to the $\pi^+\pi^-\pi^0$ invariant mass 
distribution}
\label{masssp}
In order to determine the peak cross sections for $e^+e^-$ 
annihilation into $\omega$ and $\phi$ mesons and 
the resonance parameters of excited $\omega$ states,
we fit the background-subtracted 3$\pi$ invariant-mass spectrum. 
The mass spectrum is described by the following function:
\begin{equation}
\frac{{\rm d}N}{{\rm d}m}=\sigma_{3\pi}(m)\frac{{\rm d}L}{{\rm d}m}\,R\,\varepsilon,
\label{thmspt}
\end{equation}
where $\sigma_{3\pi}(m)$ is the Born cross section for $e^+e^-\to 3\pi$,
${{\rm d}L}/{{\rm d}m}$ is the so-called ISR differential
luminosity, $\varepsilon$ is the detection efficiency as a function of mass,
and $R$ is a radiative correction factor accounting for the Born mass
spectrum distortion due to emission of several photons by the initial
electron and positron. The ISR luminosity is calculated
using the total integrated luminosity $L$ and the probability density 
function for ISR photon emission~(Eq.~(\ref{eq2})):
\begin{equation}
\frac{{\rm d}L}{{\rm d}m}=\frac{\alpha}{\pi x}\left(
(2-2x+x^2)\log\frac{1+C}{1-C}-x^2 C\right)\frac{2m}{s}\,L.
\label{ISRlum}
\end{equation}
Here $x=1-m^2/s$, $\sqrt{s}$ is the $e^+e^-$ c.m. energy, $C=\cos{\theta_0}$,
and $\theta_0$ determines the range of polar angles in the c.m. frame:
$\theta_0<\theta_\gamma<180^\circ-\theta_0$ for the ISR photon.
In our case  $\theta_0$  is equal to 20$^\circ$,
since we determine the detector efficiency using the simulation with
$20^\circ<\theta_\gamma<160^\circ$.

The Born cross section for $e^+e^-\to 3\pi$ can be written as the sum
of the contributions of four resonances:
\begin{widetext}
\begin{equation}
\sigma_{3\pi}(m)=\frac{12\pi}{m^3}F_{\rho\pi}(m)
\left|\sum_{V=\omega,\phi,\omega^\prime,\omega^{\prime\prime}}
\frac{\Gamma_V m_V^{3/2}\sqrt{{\cal B}(V\to e^+e^-){\cal B}(V\to 3\pi)}}{D_V(m)}
\frac{e^{i\phi_V}}{\sqrt{F_{\rho\pi}(m_V)}}\right|^2,
\end{equation}
\end{widetext}
where $m_V$ and $\Gamma_V$ are the mass and width of the resonance $V$,
$\phi_V$ is its phase,
${\cal B}(V\to e^+e^-)$ and ${\cal B}(V\to 3\pi)$ are the branching fractions
of $V$ into $e^+e^-$ and $3\pi$,
$$D_V(m)=m_V^2 - m^2 - i m\Gamma_V(m),\:\: \Gamma_V(m)=\sum_{f}\Gamma_f(m).$$
Here $\Gamma_f(m)$ is the mass-dependent partial width of the resonance decay
into the final state $f$, and $\Gamma_f(m_V)=\Gamma_V {\cal B}(V\to f)$.
The mass-dependent width for the $\omega$ and $\phi$ mesons has been
calculated taking into account all significant decay modes.
The corresponding formulae can be found,
for example, in Ref.~\cite{SND2003}. We assume that $V\to 3\pi$ decay proceeds
via the $\rho\pi$ intermediate state, and $F_{\rho\pi}(m)$ is the $3\pi$
phase space volume calculated under this hypothesis.
The formula for $F_{\rho\pi}$ calculation can be found in Ref.~\cite{SND2003}.

The radiative correction factor  was determined using Monte Carlo
simulation (at the generator level, with no detector simulation).
The $3\pi$ mass spectrum was generated both using only
the pure Born amplitude of the $e^+ e^- \to \pi^+ \pi^- \pi^0\gamma $
process and using a model with higher-order
radiative corrections included with the structure function method.
With the cut on the invariant mass of the $3\pi \gamma$ system,
$M_{3\pi\gamma} > 8$ GeV/$c^2$, used in our simulation,
no significant difference is found
between these two spectra.
Therefore the radiative correction factor is evaluated as the ratio
of the total cross section with $M_{3\pi\gamma}>8$ GeV/$c^2$ to the 
Born cross section and is found to be close to unity, $R=0.9994$.
The theoretical uncertainty in the radiative correction calculation 
with the structure function method does not exceed 1\%~\cite{strfun}. 
The radiative correction factor does not include 
the corrections due to leptonic and hadronic vacuum polarization.
Here we follow the generally accepted practice~\cite{vacuum} of including 
the vacuum polarization correction in the resonance electronic width.
The probability density function (PDF) for the $3\pi$ mass spectrum as 
expressed in Eq.~(\ref{thmspt}) needs to be convolved with the detector
resolution function in order to fully characterize the experimental 
mass distribution found in the data. 
\begin{figure}
\includegraphics[width=0.9\linewidth]{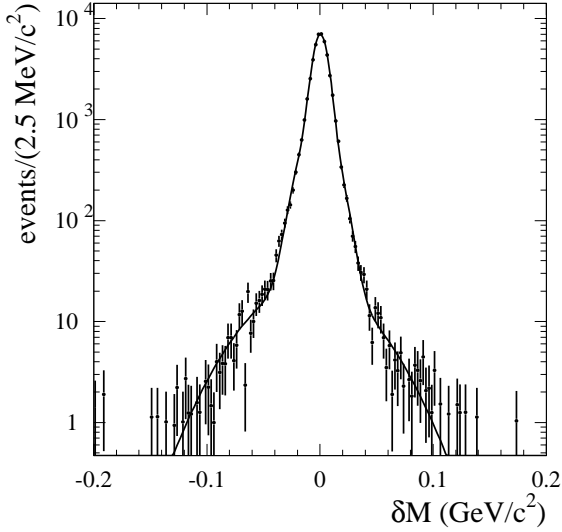}
\caption{The distribution of the difference between measured and true 
$3\pi$ mass for simulated events with $\chi^2<40$ from the mass region 
near the $\omega$. The curve is a fit to a triple-Gaussian function.}
\label{res40}
\end{figure}
The detector resolution function is
obtained using MC simulation of the detector response.
Figure~\ref{res40} shows the distribution of the difference
between measured and true $3\pi$ mass for simulated events
with $\chi^2<40$ in the mass region near the $\omega$.
For each $3\pi$ mass region,
the distribution is fit with a triple-Gaussian function.
The resolution depends on the $3\pi$ mass and is about 6, 7, and
9 MeV/$c^2$ at the $\omega$, $\phi$ and $J/\psi$ masses, respectively.
The determination of the resolution function is performed in
the $\omega$, $\phi$ and $J/\psi$ mass regions, where the available Monte Carlo
simulation statistics are high enough. For the mass region between
the $\omega$ and $\phi$ we use a linear interpolation.
Since no narrow peaks are present in the mass region between
the $\phi$ and the $J/\psi$, it is not critical to know a very detailed 
resolution function, and therefore an average resolution function is used 
over the mass range between 1.05 and 3.00 GeV.
\begin{figure}
\includegraphics[width=0.9\linewidth]{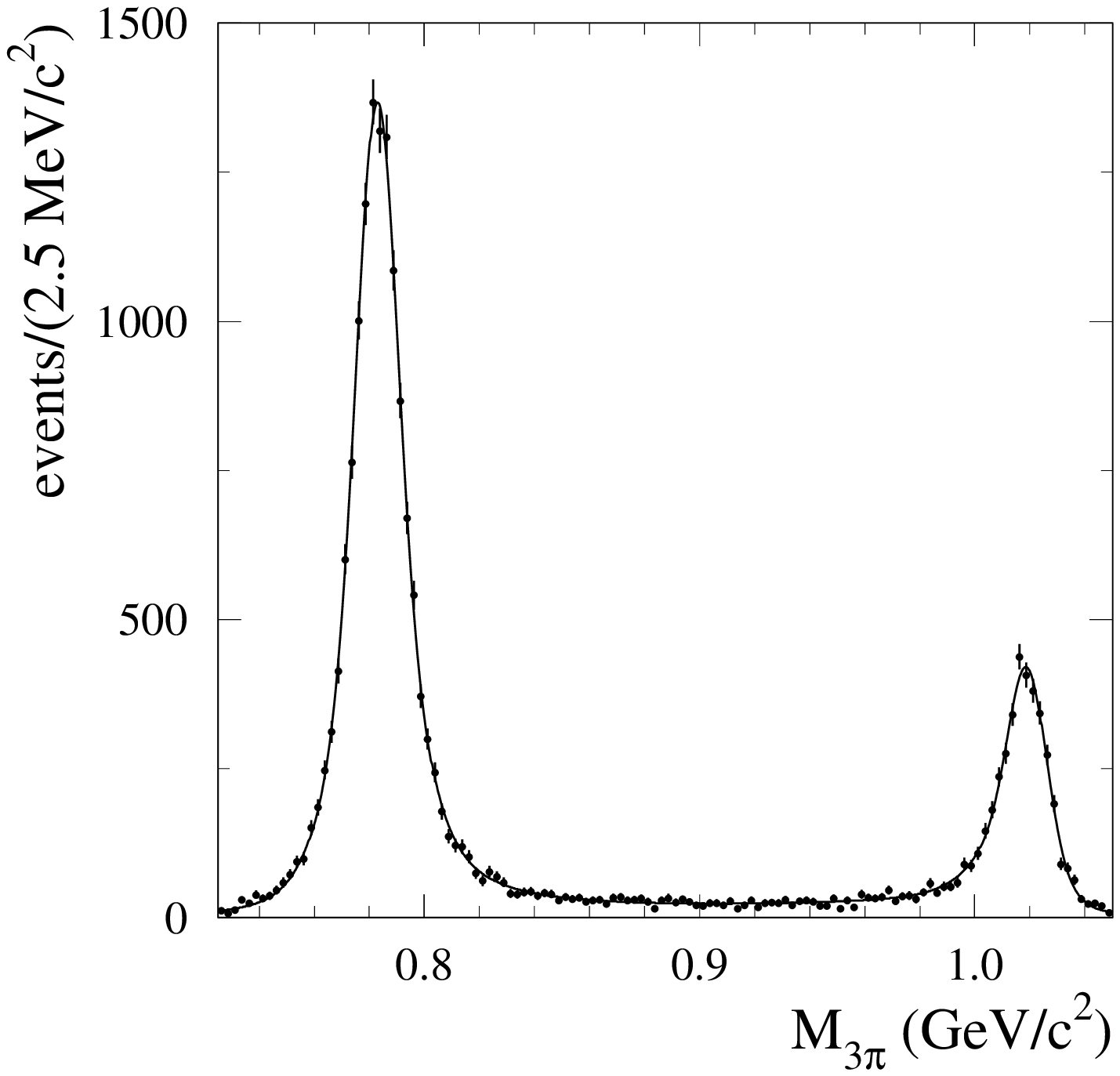}
\includegraphics[width=0.9\linewidth]{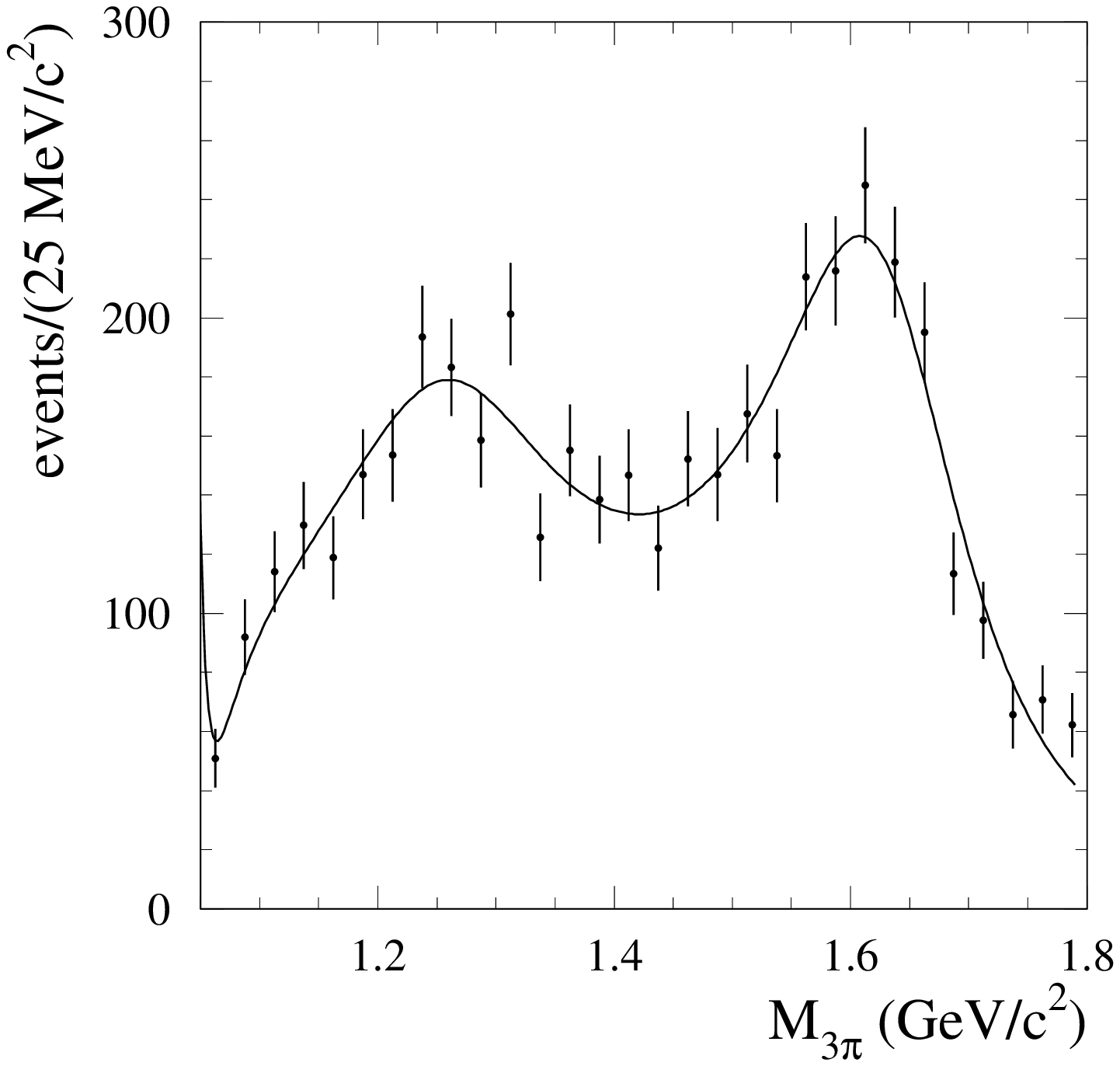}
\caption{The background-subtracted $3\pi$ mass spectrum for masses 
between 0.70 and 1.05 MeV/$c^2$
(upper plot) and for masses from 1.05 to 1.80 MeV/$c^2$ (lower plot).
The curves are
the result of the fit described in the text.}
\label{omprime_fit}
\end{figure}

A binned maximum likelihood fit is used to fit the $3\pi$ mass spectrum in
data. The bin width is chosen to be 2.5 MeV/$c^2$ for the $\omega-\phi$ mass
range and 25 MeV/$c^2$ for masses above the $\phi$.
The free parameters in the fit
are the products of the branching fractions
${\cal B}(V\to e^+e^-){\cal B}(V\to 3\pi)$, the masses $m_V$ for all four
resonances ($V=\omega,\phi,\omega^\prime,\omega^{\prime\prime}$),
and the widths for
$\omega^\prime$ and $\omega^{\prime\prime}$. The width
of the $\omega$ meson is fixed at the value of $(8.68\pm 0.13)$ MeV/$c^2$
obtained recently in the CMD2~\cite{CMD2om} and SND~\cite{SND2003}
experiments.
The width of the $\phi$ meson is fixed at
the PDG value. The relative phase between the $\omega$
and $\phi$ amplitudes, $\phi_\phi=(163\pm7)^\circ$, is taken 
from Ref.~\cite{SND2003}.
The phases of $\omega$, $\omega^\prime$, and $\omega^{\prime\prime}$
are fixed at values of $0^\circ$, $180^\circ$, and $0^\circ$~\cite{Clegg}.
Our fitting function does not take into account the contribution
of the $e^+e^-\to\omega\pi\gamma\to 3\pi\gamma$ process which proceeds
via excited $\rho$ states. In Ref.~\cite{SND2002} it is shown that  this
mechanism does not change significantly
the parameters for the $\omega^\prime$ and $\omega^{\prime\prime}$
resonances.
The fitted mass region is restricted to masses below 1.8 GeV.
To be described properly, data in the higher mass range
would require a more complicated function,
which would take into account both the resonant and the
non-resonant $3\pi$ production. 
There are no reliable models available in the literature,
and therefore our results on the
parameters of the $\omega^\prime$ and $\omega^{\prime\prime}$ states, obtained
in this fit, should only be considered a first approximation.

In order to account for a possible resolution difference between data and
simulation, the resolution function determined
from simulation is modified by adding or subtracting quadratically
an additional $\sigma_G$ to all sigmas of the triple-Gaussian 
function.
Technically, a squared sigma $\sigma^2_G$ is used as a free parameter
(with negative values allowed).
Two $\sigma_G^2$ parameters are used, one for $\omega$
and another for $\phi$ and higher masses.

The fit result is shown along with the data in Fig.~\ref{omprime_fit}.
The resulting parameters obtained from the fit ($\chi^2/\mbox{dof}=146/148$)
are the following:
\begin{eqnarray}
&{\cal B}(\omega\to e^+e^-){\cal B}(\omega\to 3\pi)=(6.70\pm0.06\pm0.27)\times 10^{-5}, \nonumber \\
&m_\omega-m_\omega^{\rm PDG}=-(0.2\pm0.1)\mbox{ MeV}/c^2,\nonumber \\
&\sigma_{G\omega}^2=(0.9\pm1.6)\mbox{ MeV}^2/{c}^4,\nonumber \\ 
&{\cal B}(\phi\to e^+e^-){\cal B}(\phi\to 3\pi)=(4.30\pm0.08\pm0.21)\times 10^{-5}, \nonumber \\
&m_\phi-m_\phi^{\rm PDG}=-(0.6\pm0.2)\mbox{ MeV}/c^2, \nonumber \\
&\sigma_{G\phi}^2=-(3.2\pm2.6)\mbox{ MeV}^2/{c}^4, \nonumber \\
&{\cal B}(\omega^\prime\to e^+e^-){\cal B}(\omega^\prime\to 3\pi)=(0.82\pm0.05\pm0.06)\times 10^{-6}, \nonumber \\
&M_{\omega^\prime}=(1350\pm20\pm20) \mbox{ MeV}/c^2, \nonumber \\
&\Gamma_{\omega^\prime}=(450\pm70\pm 70) \mbox{ MeV}/c^2, \nonumber \\
&{\cal B}(\omega^{\prime\prime}\to e^+e^-){\cal B}(\omega^{\prime\prime}\to 3\pi)=(1.3\pm0.1\pm0.1)\times 10^{-6},\nonumber \\
&M_{\omega^{\prime\prime}}=(1660\pm10\pm2)\mbox{ MeV}/c^2,\nonumber \\
&\Gamma_{\omega^{\prime\prime}}=(230\pm30\pm 20)\mbox{ MeV}/c^2.\nonumber 
\end{eqnarray}
The quoted errors correspond to the statistical and systematic uncertainties,
respectively. The systematic error for
${\cal B}(V\to e^+e^-){\cal B}(V\to 3\pi)$ includes
a statistical error from simulation, the error on the efficiency correction
(Table~\ref{effcorr}), 1.2\% uncertainty in the luminosity, 1\% theoretical
uncertainty on the radiative correction, a background subtraction uncertainty
(0.4\% at $\omega$ and 0.6\% at $\phi$), and an uncertainty arising from
errors on $\Gamma_\omega$, $\Gamma_\phi$, and $\phi_\phi$
(1\% at $\omega$ and 2.8\% at $\phi$).
The systematic errors on the masses and widths of the $\omega^\prime$
and $\omega^{\prime\prime}$ mesons are due to the background-subtraction
uncertainty and the errors on $\Gamma_\omega$, $\Gamma_\phi$, and $\phi_\phi$.

The fitted values ${\cal B}(V\to e^+e^-){\cal B}(V\to 3\pi)$ for the $\omega$ 
and $\phi$ mesons 
are in reasonable agreement with the 
corresponding 
world average values~\cite{pdg},
${\cal B}(\omega\to e^+e^-){\cal B}(\omega\to 3\pi)=(6.35\pm0.10)\times 10^{-5}$ and
${\cal B}(\phi\to e^+e^-){\cal B}(\phi\to 3\pi)=(4.52\pm0.19)\times 10^{-5}$.
The observed peak positions of both the $\omega$ and $\phi$ are shifted to 
lower masses relative to their PDG values. The shifts are about 
(0.3--0.6)$\times 10^{-3}$ of the mass values. 
The fitted values of 
$\sigma_{G}^2$ parameters have large statistical uncertainties, and lead 
to a change in the simulated resolution of (2--3)\%. 

The fitted masses and widths of the $\omega^\prime$ and
$\omega^{\prime\prime}$ mesons can be compared with the estimates
of these parameters by the PDG~\cite{pdg}:
$M_{\omega^\prime}=1400-1450\mbox{ MeV}/c^2$,
$\Gamma_{\omega^\prime}=180-250\mbox{ MeV}/c^2$,
$M_{\omega^{\prime\prime}}=1670\pm30\mbox{ MeV}/c^2$,
$\Gamma_{\omega^{\prime\prime}}=315\pm35\mbox{ MeV}/c^2$.
The PDG data are based on small data samples for
$e^+e^-\to \omega^\prime, \omega^{\prime\prime} \to 3\pi,\; \omega\pi\pi$~\cite{DM2,SND2002,CMD2opp},
$p\bar{p}\to \omega^\prime\pi^0 \to \omega \pi^0\pi^0\pi^0$~\cite{CrBar},
and
$\pi^-p\to \omega^{\prime\prime}n \to \omega\eta n$~\cite{E852} reactions.
We present a new measurement of the $\omega^\prime$ and $\omega^{\prime\prime}$
parameters based on a significantly larger data sample for the
$e^+e^-\to \omega^\prime, \omega^{\prime\prime}\to 3\pi$ reaction.
From the measured values of ${\cal B}(V\to e^+e^-){\cal B}(V\to 3\pi)$, 
the electronic widths of $\omega^\prime$ and $\omega^{\prime\prime}$
can be estimated. Assuming
that ${\cal B}(\omega^\prime\to 3\pi)\approx 1$ and
${\cal B}(\omega^{\prime\prime}\to 3\pi)\approx 0.5$ 
we derive that
$\Gamma(\omega^\prime\to e^+e^-)\approx 370$ eV and 
$\Gamma(\omega^{\prime\prime}\to e^+e^-)\approx 570$ eV.
The large values of these widths, comparable with 
$\Gamma(\omega\to e^+e^-)\approx 600$ eV, are in disagreement with 
expectations of the quark model, which predicts 
at least one order of magnitude lower 
values for the electronic widths for the excited meson states (see, for 
example, Ref.~\cite{Isgur}).

\begin{figure*}[t]
\includegraphics[width=0.9\linewidth]{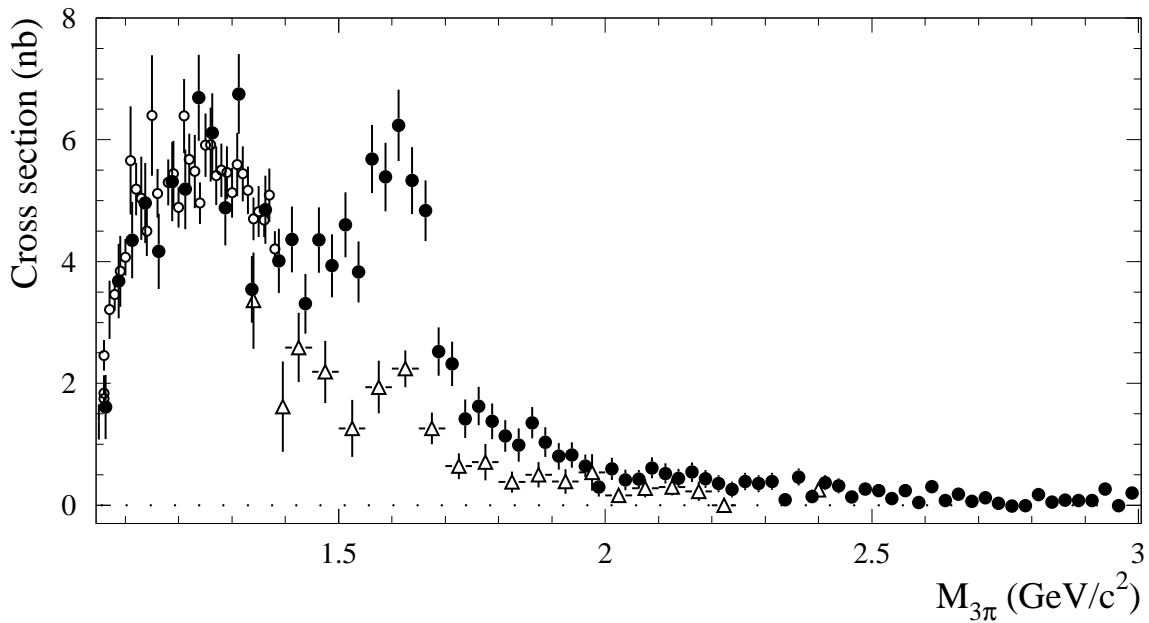}
\caption{The $e^+e^-\to \pi^+\pi^-\pi^0$ cross section measured in
this work (filled circles), by SND (open circles), and DM2 (open triangles).}
\label{cscomp}
\end{figure*}
\section{\boldmath Measurement of the $e^+e^-\to\pi^+\pi^-\pi^0$ 
cross section}
\begin{table*}
\caption{ $e^+e^-$ c.m. energy ($\sqrt{s^\prime}$), number of selected events
after $3\pi$-mass-resolution correction ($N_{corr}$), detection efficiency
($\varepsilon$), differential ISR luminosity ($L$), and measured cross section
($\sigma$) for $e^+e^-\to \pi^+\pi^-\pi^0$. All values are calculated for 25
MeV bin size. The quoted errors are statistical and systematic.}
\label{sumtab}
\begin{ruledtabular}
\begin{tabular}{cccccccccc}
$\sqrt{s^\prime}$&  $N_{corr}$     &$\varepsilon$  &     $L$   &  $\sigma$&$\sqrt{s^\prime}$&  $N_{corr}$     &$\varepsilon$  & $L$       &  $\sigma$\\
  (GeV)     &                 &      (\%)       &(nb$^{-1}$)&  (nb)      &  (GeV)     &                 &      (\%)       &(nb$^{-1}$)&  (nb)     \\
\hline
1.0625&$ 40\pm 13\pm 3$&$  9.8\pm  0.4$& 254 &$ 1.61\pm 0.53\pm 0.13$& 2.0375&$ 22\pm  9\pm 2$&$ 10.2\pm  0.5$& 511 &$ 0.41\pm 0.17\pm 0.04$\\
1.0875&$ 94\pm 16\pm 4$&$  9.8\pm  0.4$& 260 &$ 3.68\pm 0.61\pm 0.21$& 2.0625&$ 23\pm  8\pm 1$&$ 10.2\pm  0.5$& 518 &$ 0.43\pm 0.15\pm 0.02$\\
1.1125&$114\pm 16\pm 3$&$  9.8\pm  0.4$& 266 &$ 4.35\pm 0.63\pm 0.21$& 2.0875&$ 33\pm  9\pm 1$&$ 10.2\pm  0.5$& 525 &$ 0.61\pm 0.17\pm 0.04$\\
1.1375&$133\pm 18\pm 4$&$  9.9\pm  0.4$& 272 &$ 4.96\pm 0.66\pm 0.25$& 2.1125&$ 28\pm  9\pm 1$&$ 10.2\pm  0.5$& 532 &$ 0.52\pm 0.16\pm 0.03$\\
1.1625&$115\pm 17\pm 4$&$  9.9\pm  0.4$& 278 &$ 4.17\pm 0.62\pm 0.22$& 2.1375&$ 24\pm  8\pm 1$&$ 10.2\pm  0.5$& 539 &$ 0.44\pm 0.15\pm 0.03$\\
1.1875&$150\pm 18\pm 4$&$  9.9\pm  0.4$& 285 &$ 5.31\pm 0.65\pm 0.25$& 2.1625&$ 30\pm  9\pm 1$&$ 10.2\pm  0.5$& 547 &$ 0.54\pm 0.16\pm 0.03$\\
1.2125&$150\pm 19\pm 4$&$  9.9\pm  0.4$& 291 &$ 5.19\pm 0.65\pm 0.26$& 2.1875&$ 24\pm  8\pm 1$&$ 10.2\pm  0.5$& 554 &$ 0.43\pm 0.15\pm 0.03$\\
1.2375&$198\pm 21\pm 5$&$  9.9\pm  0.4$& 297 &$ 6.69\pm 0.71\pm 0.31$& 2.2125&$ 20\pm  8\pm 1$&$ 10.2\pm  0.5$& 561 &$ 0.36\pm 0.14\pm 0.02$\\
1.2625&$185\pm 20\pm 4$&$ 10.0\pm  0.4$& 304 &$ 6.11\pm 0.65\pm 0.27$& 2.2375&$ 15\pm  8\pm 1$&$ 10.2\pm  0.5$& 569 &$ 0.26\pm 0.13\pm 0.02$\\
1.2875&$151\pm 19\pm 5$&$ 10.0\pm  0.4$& 310 &$ 4.88\pm 0.62\pm 0.25$& 2.2625&$ 23\pm  8\pm 1$&$ 10.2\pm  0.5$& 576 &$ 0.39\pm 0.14\pm 0.03$\\
1.3125&$214\pm 21\pm 4$&$ 10.0\pm  0.4$& 316 &$ 6.75\pm 0.65\pm 0.31$& 2.2875&$ 21\pm  8\pm 1$&$ 10.2\pm  0.5$& 584 &$ 0.36\pm 0.14\pm 0.02$\\
1.3375&$115\pm 18\pm 6$&$ 10.0\pm  0.4$& 323 &$ 3.54\pm 0.55\pm 0.23$& 2.3125&$ 23\pm  8\pm 1$&$ 10.2\pm  0.5$& 591 &$ 0.39\pm 0.14\pm 0.03$\\
1.3625&$160\pm 19\pm 4$&$ 10.0\pm  0.4$& 329 &$ 4.85\pm 0.56\pm 0.24$& 2.3375&$  5\pm  7\pm 1$&$ 10.1\pm  0.5$& 599 &$ 0.09\pm 0.11\pm 0.01$\\
1.3875&$136\pm 18\pm 4$&$ 10.1\pm  0.4$& 335 &$ 4.01\pm 0.53\pm 0.21$& 2.3625&$ 28\pm  9\pm 1$&$ 10.1\pm  0.6$& 606 &$ 0.46\pm 0.14\pm 0.03$\\
1.4125&$150\pm 19\pm 5$&$ 10.1\pm  0.4$& 342 &$ 4.36\pm 0.54\pm 0.24$& 2.3875&$  9\pm  7\pm 1$&$ 10.1\pm  0.6$& 614 &$ 0.15\pm 0.11\pm 0.01$\\
1.4375&$116\pm 17\pm 5$&$ 10.1\pm  0.4$& 348 &$ 3.31\pm 0.49\pm 0.20$& 2.4125&$ 23\pm  8\pm 1$&$ 10.1\pm  0.6$& 621 &$ 0.37\pm 0.12\pm 0.02$\\
1.4625&$156\pm 19\pm 4$&$ 10.1\pm  0.4$& 355 &$ 4.36\pm 0.54\pm 0.21$& 2.4375&$ 20\pm  8\pm 1$&$ 10.1\pm  0.6$& 629 &$ 0.32\pm 0.12\pm 0.02$\\
1.4875&$144\pm 19\pm 3$&$ 10.1\pm  0.4$& 361 &$ 3.93\pm 0.52\pm 0.20$& 2.4625&$  9\pm  6\pm 1$&$ 10.1\pm  0.6$& 637 &$ 0.14\pm 0.10\pm 0.01$\\
1.5125&$172\pm 20\pm 4$&$ 10.1\pm  0.4$& 368 &$ 4.60\pm 0.53\pm 0.22$& 2.4875&$ 17\pm  7\pm 1$&$ 10.0\pm  0.6$& 645 &$ 0.26\pm 0.11\pm 0.02$\\
1.5375&$146\pm 19\pm 3$&$ 10.2\pm  0.4$& 374 &$ 3.83\pm 0.50\pm 0.19$& 2.5125&$ 16\pm  7\pm 1$&$ 10.0\pm  0.6$& 653 &$ 0.24\pm 0.11\pm 0.02$\\
1.5625&$220\pm 22\pm 5$&$ 10.2\pm  0.4$& 381 &$ 5.69\pm 0.56\pm 0.28$& 2.5375&$  7\pm  6\pm 1$&$ 10.0\pm  0.6$& 660 &$ 0.11\pm 0.09\pm 0.01$\\
1.5875&$213\pm 22\pm 5$&$ 10.2\pm  0.4$& 387 &$ 5.39\pm 0.56\pm 0.27$& 2.5625&$ 16\pm  7\pm 1$&$ 10.0\pm  0.6$& 668 &$ 0.24\pm 0.10\pm 0.02$\\
1.6125&$250\pm 23\pm 5$&$ 10.2\pm  0.5$& 394 &$ 6.24\pm 0.58\pm 0.31$& 2.5875&$  3\pm  6\pm 1$&$ 10.0\pm  0.6$& 676 &$ 0.05\pm 0.08\pm 0.01$\\
1.6375&$218\pm 22\pm 5$&$ 10.2\pm  0.5$& 401 &$ 5.33\pm 0.55\pm 0.27$& 2.6125&$ 21\pm  7\pm 1$&$  9.9\pm  0.6$& 684 &$ 0.30\pm 0.10\pm 0.02$\\
1.6625&$201\pm 21\pm 4$&$ 10.2\pm  0.5$& 407 &$ 4.84\pm 0.50\pm 0.24$& 2.6375&$  6\pm  6\pm 1$&$  9.9\pm  0.6$& 693 &$ 0.08\pm 0.08\pm 0.01$\\
1.6875&$107\pm 17\pm 3$&$ 10.2\pm  0.5$& 414 &$ 2.52\pm 0.40\pm 0.14$& 2.6625&$ 13\pm  7\pm 1$&$  9.9\pm  0.6$& 701 &$ 0.18\pm 0.10\pm 0.01$\\
1.7125&$100\pm 16\pm 2$&$ 10.2\pm  0.5$& 421 &$ 2.32\pm 0.36\pm 0.12$& 2.6875&$  5\pm  7\pm 1$&$  9.9\pm  0.6$& 709 &$ 0.07\pm 0.09\pm 0.01$\\
1.7375&$ 62\pm 14\pm 2$&$ 10.2\pm  0.5$& 427 &$ 1.42\pm 0.31\pm 0.08$& 2.7125&$  9\pm  6\pm 1$&$  9.8\pm  0.6$& 717 &$ 0.12\pm 0.09\pm 0.01$\\
1.7625&$ 72\pm 14\pm 2$&$ 10.2\pm  0.5$& 434 &$ 1.63\pm 0.31\pm 0.09$& 2.7375&$  2\pm  5\pm 1$&$  9.8\pm  0.6$& 725 &$ 0.03\pm 0.08\pm 0.01$\\
1.7875&$ 62\pm 13\pm 2$&$ 10.2\pm  0.5$& 441 &$ 1.38\pm 0.29\pm 0.07$& 2.7625&$ -1\pm  5\pm 1$&$  9.8\pm  0.6$& 734 &$-0.01\pm 0.07\pm 0.01$\\
1.8125&$ 52\pm 12\pm 1$&$ 10.2\pm  0.5$& 448 &$ 1.14\pm 0.26\pm 0.06$& 2.7875&$  0\pm  5\pm 1$&$  9.8\pm  0.6$& 742 &$ 0.00\pm 0.07\pm 0.01$\\
1.8375&$ 46\pm 13\pm 3$&$ 10.2\pm  0.5$& 455 &$ 0.99\pm 0.27\pm 0.07$& 2.8125&$ 13\pm  7\pm 1$&$  9.7\pm  0.6$& 751 &$ 0.17\pm 0.09\pm 0.01$\\
1.8625&$ 64\pm 12\pm 2$&$ 10.3\pm  0.5$& 462 &$ 1.35\pm 0.26\pm 0.07$& 2.8375&$  4\pm  6\pm 1$&$  9.7\pm  0.6$& 759 &$ 0.05\pm 0.08\pm 0.01$\\
1.8875&$ 50\pm 12\pm 2$&$ 10.3\pm  0.5$& 468 &$ 1.04\pm 0.24\pm 0.07$& 2.8625&$  6\pm  5\pm 1$&$  9.7\pm  0.6$& 768 &$ 0.08\pm 0.07\pm 0.01$\\
1.9125&$ 39\pm 11\pm 2$&$ 10.3\pm  0.5$& 475 &$ 0.80\pm 0.22\pm 0.06$& 2.8875&$  6\pm  5\pm 1$&$  9.6\pm  0.6$& 776 &$ 0.08\pm 0.07\pm 0.01$\\
1.9375&$ 41\pm 10\pm 1$&$ 10.3\pm  0.5$& 482 &$ 0.83\pm 0.21\pm 0.05$& 2.9125&$  6\pm  5\pm 1$&$  9.6\pm  0.6$& 785 &$ 0.08\pm 0.07\pm 0.01$\\
1.9625&$ 32\pm  9\pm 1$&$ 10.3\pm  0.5$& 489 &$ 0.64\pm 0.19\pm 0.04$& 2.9375&$ 20\pm  7\pm 1$&$  9.6\pm  0.6$& 794 &$ 0.26\pm 0.09\pm 0.02$\\
1.9875&$ 15\pm  8\pm 1$&$ 10.3\pm  0.5$& 496 &$ 0.30\pm 0.16\pm 0.03$& 2.9625&$ -1\pm  5\pm 1$&$  9.5\pm  0.6$& 802 &$-0.01\pm 0.06\pm 0.01$\\
2.0125&$ 31\pm  9\pm 1$&$ 10.2\pm  0.5$& 503 &$ 0.60\pm 0.18\pm 0.04$& 2.9875&$ 16\pm  6\pm 1$&$  9.5\pm  0.6$& 811 &$ 0.20\pm 0.08\pm 0.01$\\
\end{tabular}
\end{ruledtabular}
\end{table*}
The cross section for $e^+e^-\to\pi^+\pi^-\pi^0$,
in the energy ($\sqrt{s^\prime}$) range between 1.05 and 3.00 GeV,
is calculated from the $3\pi$ mass spectrum using
\begin{equation}
\sigma_{3\pi}(m)=\frac{({\rm d}N/{\rm d}m)_{corr}}{\varepsilon\, R\, 
{\rm d}L/{\rm d}m}, 
\end{equation}
where $m\equiv \sqrt{s^\prime}$ is the $3\pi$ invariant mass,
and $({\rm d}N/{\rm d}m)_{corr}$ is the mass spectrum corrected for resolution
effects.

The resolution-corrected mass spectrum is obtained by first
subtracting the events with an actual $3\pi$ invariant mass outside
the 1.05--3.00 GeV/$c^2$ region (tails of the $\phi$ and $J/\psi$ mass
distribution). The number of $\phi$-meson
events with measured mass above 1.05 GeV/$c^2$ is estimated from simulation.
We subtract $10\pm5$ events from the first mass bin (1.05--1.075 GeV/$c^2$).
The number of $J/\psi$ events contributing to the mass region under
study is found to be $1\pm 1$.
Second, the detector resolution is deconvolved by using a
migration matrix $A$ that gives the probability that an event with
true mass in bin $j$ is actually reconstructed in bin $i$:
\begin{equation}
\left( \frac{{\rm d}N}{{\rm d}m}\right)^{rec}_i=
\sum_j A_{ij}\left(\frac{{\rm d}N}{{\rm d}m}\right)^{true}_j.
\end{equation}
The inverse of this migration matrix ($A^{-1}_{ij}$) is then
applied to the measured spectrum.
In our case where the spectrum has no narrow structures and
the smearing is small, we can neglect the mass dependence of the resolution
and use the average resolution function to build the migration matrix.
This allows a determination of the inverse matrix that is robust
against statistical fluctuations.
The resolution function is obtained from simulation and takes into account
our background subtraction procedure.
In practice, the only elements of the
migration matrix that are significantly different from zero are the
diagonal elements, which are around 0.84, and the elements next to the
diagonal, which are around 0.08.
The resolution correction procedure changes insignificantly
the shape of the mass distribution but leads to
an increase in the errors (by $\approx $20\%)
and their correlation. The number of events in each bin changes by no more
than half its statistical uncertainty. In particular, near 3$\pi$
mass of 1.6~GeV this change is less than 3\%.
The number of events for each mass bin is listed in Table~\ref{sumtab}.
The quoted errors correspond to the statistical
(including the contribution of the background subtraction
for the $e^+e^-\to K^+K^-\pi^0\gamma$ and
$e^+e^-\to\pi^+\pi^-\pi^0\pi^0$ processes) and systematic
(due to the $\alpha$ and $\beta$ parameters) uncertainties.

Table~\ref{sumtab} also contains the values of the detection efficiency
and the ISR differential luminosity calculated according to Eq.~(\ref{ISRlum}).
The errors on the detection efficiency include the uncertainties on
the efficiency correction and the statistical errors from the simulation.

The calculated cross section is shown in Fig.~\ref{cscomp} and listed in
Table~\ref{sumtab}. The quoted errors are statistical and systematic.
The latter includes the systematic error contributions from 
the number of events, and uncertainties in the detection efficiencies
and in the calculation of the radiative correction.
Note that the systematic errors for different mass 
bins are fully correlated.

Most of experiments at low energy measure so-called ``dressed'' cross
sections (see, for example, Ref.~\cite{SND2002}), which include
the vacuum polarization corrections. Since our radiative correction
factor $R$ does not take into account vacuum polarization we
also measure the ``dressed'' cross section.
A comparison of our measurements with other
$e^+e^-$ data is shown in Fig.~\ref{cscomp}. Our values are in good  
agreement
with SND measurements, but significantly exceed DM2 results.
\section{\boldmath Measurement of the $J/\psi\to 3\pi$ branching fraction}\label{jpsi}
\begin{figure}
\includegraphics[width=0.9\linewidth]{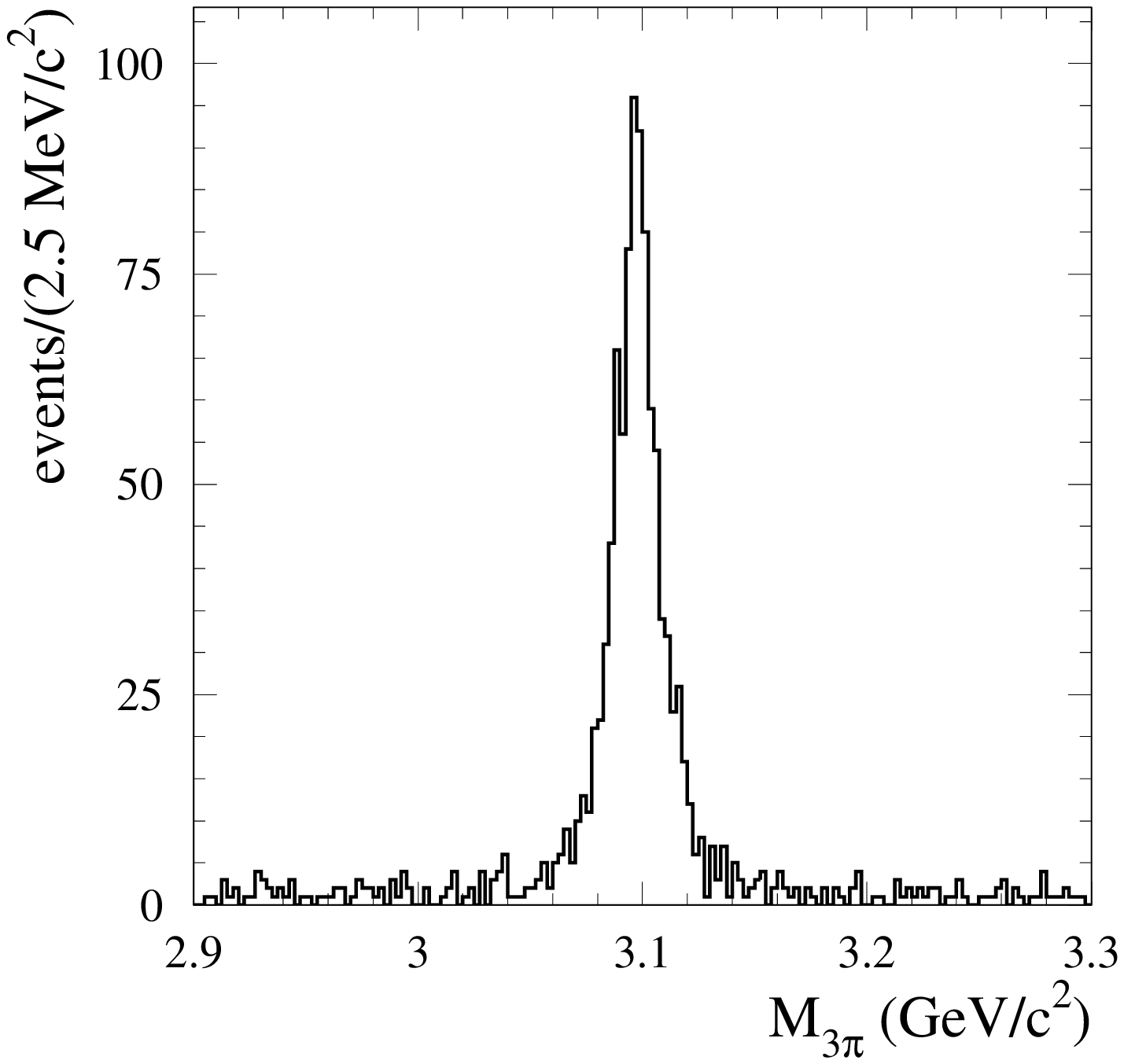}
\caption{$3\pi$ mass spectrum for selected 
$e^+e^-\to J/\psi\gamma\to \pi^+\pi^-\pi^0\gamma$ events.}
\label{psiexp}
\end{figure}
\begin{table}[t]
\begin{ruledtabular}
\caption{$N_{signal}$ and $N_{side}$ are the numbers of selected events
in the signal region ($3.0\leq M_{3\pi}\leq 3.2$ GeV/$c^2$) and the sidebands
($2.9\leq M_{3\pi}<3.0$ and $3.2\leq M_{3\pi}<3.3$ GeV/$c^2$), respectively.}
\begin{tabular}{cccc}
        & $N_{signal}$ & $N_{side}$ & $N_{signal}-N_{side}$\\
data  & 1023  & 103 & 920$\pm$34   \\
  MC  & 1825  & 13  & 1812$\pm$43   \\
\end{tabular}
\end{ruledtabular}
\label{jpsitab}
\end{table}
The $3\pi$ mass spectrum for selected events in the $J/\psi$ mass region
is shown in Fig.~\ref{psiexp}. The small width of the $J/\psi$ resonance
leads to negligible peaking background. For example,
$e^+e^-\to J/\psi\gamma\to K^+K^-\pi^0\gamma$ events reconstructed
under the $3\pi\gamma$ hypothesis have a $3\pi$ invariant mass in the range
2.8 to 3.0 GeV/$c^2$. Since the nonresonant background is small and
well described by a linear function,
a mass-sideband subtraction method is used to
determine the number of $J/\psi$ events.
Table~\ref{jpsitab} shows the numbers of data and simulated 3$\pi\gamma$
events in the signal region ($3.0<M_{3\pi}<3.2$ GeV/$c^2$) and in the sidebands
($2.9<M_{3\pi}<3.0$ GeV/$c^2$ and $3.2<M_{3\pi}<3.3$ GeV/$c^2$).

The Monte Carlo simulation of the number of events in the signal and
sideband regions is used to estimate a detection efficiency of
$\varepsilon_{MC}$ = 0.101$\pm$0.002.
The data-MC simulation differences discussed earlier are used to correct
the former efficiency value by $(11\pm 7)\%$.
\begin{figure}
\includegraphics[width=0.94\linewidth]{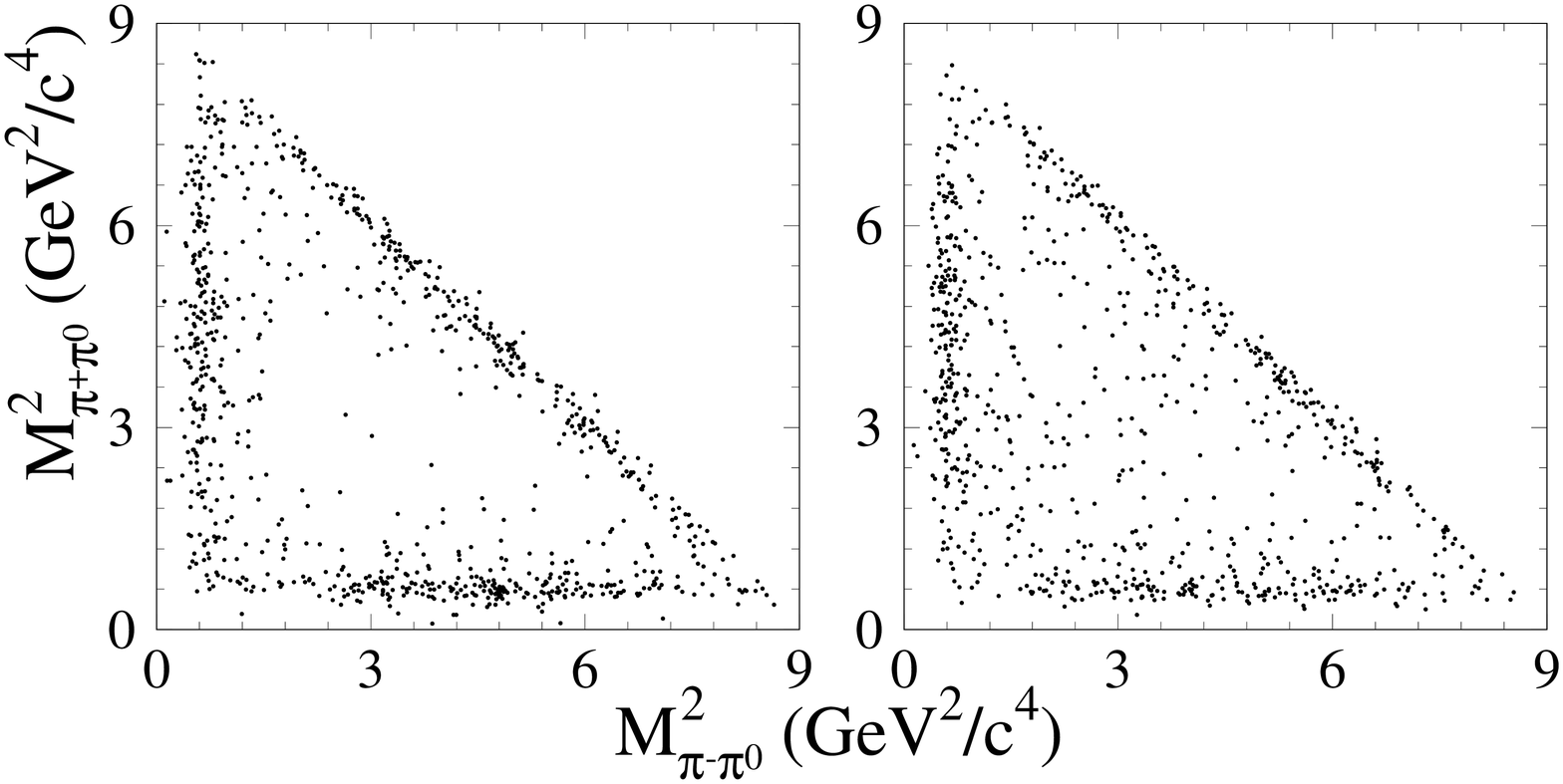}
\caption{The Dalitz plot for $J/\psi\to 3\pi$ candidates
in data (left) and simulation (right).}
\label{dalitz_psi}
\end{figure}

The simulation uses the $\rho\pi$ model of $J/\psi\to 3\pi$ decay.
In order to check the model dependence of the detection efficiency,
the Dalitz plot for events in the $J/\psi$ peak
(Fig.~\ref{dalitz_psi}) is analyzed.
It is seen that the main mechanism for $J/\psi\to 3\pi$ decay is $\rho\pi$.
There is, however, a difference between the data and simulated plots
(an absence of events in the center of the Dalitz plot for data),
which can be a manifestation of negative interference with
the contribution of intermediate states other than the $\rho\pi$~\cite{Dalitz}.
The influence of this difference on the detection efficiency
is studied by excluding events located in the center of the Dalitz plot in
the simulated sample, and recomputing the detection efficiency.
The result is a (1.1$\pm$0.6)\% rise in efficiency. This correction is
included with a systematic error of 1.1\% in the final
calculation of the detection efficiency, which is determined to be
$0.092\pm0.006$.

The cross section for
$e^+e^-\to J/\psi\gamma\to \pi^+\pi^-\pi^0\gamma$ for
$20^\circ<\theta_\gamma<160^\circ$ 
is calculated as
$$\sigma(20^\circ<\theta_\gamma<160^\circ)=\frac{N_{signal}-N_{side}}
{\varepsilon\, R\, L}=(112\pm4\pm8)\mbox{ fb}.$$
The radiative-correction factor $R=\sigma/\sigma_{Born}$, of
$1.005\pm0.002\pm0.010$ used here,
is obtained from a MC simulation at the generator level 
(no detector simulation).
The total integrated luminosity for the data sample  is 
$(89.3\pm 1.1)$ fb$^{-1}$.
From the measured cross section and Eq.~(\ref{eq3}),
the following product can be determined:
\begin{eqnarray}
\lefteqn
{\Gamma(J/\psi\to e^+e^-){\cal B}(J/\psi\to 3\pi)}\nonumber\\
& &=(0.122\pm0.005\pm0.008)\mbox{ keV}.\nonumber
\end{eqnarray}
The systematic error includes the uncertainties on the detection efficiency,
the integrated luminosity, and the radiative correction.

The most precise measurement of the electronic width was made in the analysis
of $e^+e^-\to J/\psi\gamma\to \mu^+\mu^-\gamma$
by \babar~\cite{BAD602}:
$\Gamma(J/\psi\to e^+e^-)=(5.61\pm0.20)$ keV.
Using the latter measurement, the $J/\psi\to 3\pi$ branching
fraction is calculated to be
$${\cal B}(J/\psi\to 3\pi)=(2.18\pm0.19)\%,$$
which is in substantial disagreement ($\sim 3\sigma$) with
the world average value (see Sec.~\ref{intro}) of $(1.47\pm0.13)\%$, 
but agrees with the result from the BES collaboration~\cite{BES3pi}:
${\cal B}(J/\psi\to 3\pi)=(2.10\pm0.12)\%$.

\section{Summary}
The process $e^+e^-\to\pi^+\pi^-\pi^0\gamma$ was studied
for the $3\pi$ invariant masses up to 3 GeV/$c^2$ and at the $J/\psi$ mass.
From the measured $3\pi$ mass spectrum we obtained
the $e^+e^-\to \pi^+\pi^-\pi^0$ cross section for the $1.05<\sqrt{s^\prime}<3$
GeV energy range. The results are in agreement with the SND
measurement~\cite{SND2001} for $\sqrt{s^\prime}<1.4$ and significantly
exceed the DM2 data~\cite{DM2} in the $1.4<\sqrt{s^\prime}<2.2$ range.
The $e^+e^-\to \pi^+\pi^-\pi^0$ cross section  in the energy range up to
1.8 GeV is described well by a sum of the contributions of four isoscalar
resonances: $\omega$, $\phi$, $\omega^\prime$, and $\omega^{\prime\prime}$.
From the fit of the $3\pi$ mass spectrum we obtained the following parameters
for these resonances:
\begin{eqnarray}
&{\cal B}(\omega\to e^+e^-){\cal B}(\omega\to 3\pi)=(6.70\pm0.06\pm0.27)\times 10^{-5}, \nonumber \\
&{\cal B}(\phi\to e^+e^-){\cal B}(\phi\to 3\pi)=(4.30\pm0.08\pm0.21)\times 10^{-5}, \nonumber \\
&{\cal B}(\omega^\prime\to e^+e^-){\cal B}(\omega^\prime\to 3\pi)=(0.82\pm0.05\pm0.06)\times 10^{-6}, \nonumber \\                 
&M_{\omega^\prime}=(1350\pm20\pm20)\mbox{ MeV}/c^2, \nonumber \\                                                    
&\Gamma_{\omega^\prime}=(450\pm70\pm70)\mbox{ MeV}/c^2, \nonumber \\                                                
&{\cal B}(\omega^{\prime\prime}\to e^+e^-){\cal B}(\omega^{\prime\prime}\to 3\pi)=(1.3\pm0.1\pm0.1)\times 10^{-6},\nonumber \\    
&M_{\omega^{\prime\prime}}=(1660\pm10\pm2)\mbox{ MeV}/c^2,\nonumber \\                                             
&\Gamma_{\omega^{\prime\prime}}=(230\pm30\pm20)\mbox{ MeV}/c^2.\nonumber                                            
\end{eqnarray}
The electronic widths of $\omega^\prime$ and
$\omega^{\prime\prime}$ corresponding to these resonance parameters,
$\Gamma(\omega^\prime\to e^+e^-)\approx 370$ eV and
$\Gamma(\omega^{\prime\prime}\to e^+e^-)\approx 570$ eV, are comparable with
the $\omega(782)$ electronic width, in disagreement with expectations
of the quark model (see, for example, Ref.~\cite{Isgur}).

From the measured number of events in the
$e^+e^-\to J/\psi\gamma\to\pi^+\pi^-\pi^0\gamma$ reaction we
determine
\begin{eqnarray}
\lefteqn
{\Gamma(J/\psi\to e^+e^-){\cal B}(J/\psi\to 3\pi)}\nonumber\\
& & =(0.122\pm0.005\pm0.008)\mbox{ keV}.\nonumber
\end{eqnarray}
Dividing this value by 
$\Gamma(J/\psi\to e^+e^-)=(5.61\pm0.20)$ keV~\cite{BAD602}
we obtain
${\cal B}(J/\psi\to 3\pi)=(2.18\pm0.19)\%$,
which is in $\sim$3$\sigma$ disagreement with
the world average value of $(1.47\pm0.13)\%$
(see Sec.~\ref{intro}), but agrees with the recent
result from the BES Collaboration~\cite{BES3pi}: $(2.10\pm 0.12)\%$.

\section{ \boldmath Acknowledgments}
\label{acknowl}
We are grateful for the 
extraordinary contributions of our \pep2\ colleagues in
achieving the excellent luminosity and machine conditions
that have made this work possible.
The success of this project also relies critically on the 
expertise and dedication of the computing organizations that 
support \babar.
The collaborating institutions wish to thank 
SLAC for its support and the kind hospitality extended to them. 
This work is supported by the
US Department of Energy
and National Science Foundation, the
Natural Sciences and Engineering Research Council (Canada),
Institute of High Energy Physics (China), the
Commissariat \`a l'Energie Atomique and
Institut National de Physique Nucl\'eaire et de Physique des Particules
(France), the
Bundesministerium f\"ur Bildung und Forschung and
Deutsche Forschungsgemeinschaft
(Germany), the
Istituto Nazionale di Fisica Nucleare (Italy),
the Foundation for Fundamental Research on Matter (The Netherlands),
the Research Council of Norway, the
Ministry of Science and Technology of the Russian Federation, and the
Particle Physics and Astronomy Research Council (United Kingdom). 
Individuals have received support from 
CONACyT (Mexico),
the A. P. Sloan Foundation, 
the Research Corporation,
and the Alexander von Humboldt Foundation.


\begin{thebibliography}{99}

\bibitem{arbus} A.B.~Arbuzov {\em et al.}, 
JHEP {\bf 9812}, 009 (1998). 
\bibitem{kuhn} S.~Binner, J.H.~Kuhn, and K.~Melnikov, 
Phys. Lett. B {\bf 459}, 279 (1999).
\bibitem{ivanch} M.~Benayoun {\em et al.}, 
Mod. Phys. Lett. A {\bf 14}, 2605 (1999).

\bibitem{KLOE} KLOE Collaboration, A.~Aloisio {\em et al.}, 
submitted to Phys. Lett. B, hep-ex/0407048.

\bibitem{pdg} Review of Particle Physics,
S.~Eidelman {\em et al.}, Phys. Lett. B {\bf 592}, 1 (2004).

\bibitem{SND2002} SND Collaboration, M.N.~Achasov {\em et al.},
Phys. Rev. D {\bf 66}, 032001 (2002).
\bibitem{DM2} DM2 Collaboration, A.~Antonelli {\em et al.},
Z. Phys. C {\bf 56}, 15 (1992). 

\bibitem{BAD602} \babar\ Collaboration, B.~Aubert {\em et al.},
Phys. Rev. D {\bf 69}, 011103 (2004). 

\bibitem{JEAN}B.~Jean-Marie {\em et al.},
Phys. Rev. Lett. {\bf 36}, 291 (1976).

\bibitem{BARTEL}W.~Bartel {\em et al.},
Phys. Lett. B {\bf 64}, 483 (1976).

\bibitem{BRANDELIK}DASP Collaboration, R.~Brandelik {\em et al.},
Phys. Lett. B {\bf 74}, 292 (1978).

\bibitem{ALEXANDER}PLUTO Collaboration, G.~Alexander {\em et al.},
Phys. Lett. B {\bf 72}, 493 (1978).

\bibitem{FRANKLIN}M.E.B.~Franklin {\em et al.}, 
Phys. Rev. Lett. {\bf 51}, 963 (1983).

\bibitem{COFFMAN} MARK-III Collaboration, D.~Coffman {\em et al.},
Phys. Rev. D {\bf 38}, 2695 (1988).

\bibitem{BAI}BES Collaboration, J.Z.~Bai {\em et al.},
Phys. Rev. D {\bf 54}, 1221 (1996).

\bibitem{BES3pi}BES Collaboration, J.Z.~Bai {\em et al.},
Phys. Rev. D {\bf 70}, 012005 (2004).

\bibitem{ref:babar-nim} \babar\ Collaboration, B.~Aubert {\em et al.},
Nucl. Instr. and Meth. A {\bf 479}, 1 (2002).

\bibitem{ckhhad} H.~Czyz and J.H.~Kuhn, 
Eur. Phys. J. C {\bf 18}, 497 (2001).

\bibitem{strfun} M.~Caffo, H.~Czyz, and E.~Remiddi, 
Nuo. Cim. {\bf 110A}, 515 (1997);
Phys. Lett. B {\bf 327}, 369 (1994). 

\bibitem{PHOTOS}
E.~Barberio and Z.~Was, Comput. Phys. Commun. {\bf 79}, 291 (1994).

\bibitem{Phokhara} G.~Rodrigo, H.~Czyz, J.H.~Kuhn, and M.~Szopa,
Eur. Phys. J. C {\bf 24}, 71 (2002).

\bibitem{Jetset} T.~Sjostrand, 
Comput. Phys. Commun. {\bf 82}, 74 (1994).

\bibitem{KORALB} 
S.~Jadach and Z.~Was,
Comput. Phys. Commun. {\bf 85}, 453 (1995).

\bibitem{ref:geant4} S.~Agostinelli {\em et al.},
Nucl. Instr. and Meth. A {\bf 506}, 250 (2003).

\bibitem{SND2003} SND Collaboration, M.N.~Achasov {\em et al.},
Phys. Rev. D {\bf 68}, 052006 (2003).

\bibitem{SND2001} SND Collaboration, M.N.~Achasov {\em et al.},
Phys. Rev. D {\bf 65}, 032002 (2002).

\bibitem{DAFNE} KLOE Collaboration, A.~Aloisio {\em et al.},
Phys. Lett. B {\bf 561}, 55 (2003).

\bibitem{vacuum} Crystal Ball Collaboration,                                        
Z. Jakubowski {\em et al.}, Z.\ Phys.\ C {\bf 40}, 49 (1988).

\bibitem{CMD2om} CMD-2 Collaboration, R.R.~Akhmetshin {\em et al.},
Phys. Lett. B {\bf 476}, 33 (2000); the results were corrected in 
Phys. Lett. B {\bf 578}, 285 (2004).

\bibitem{Clegg} A.B.~Clegg and A.~Donnachie,
Z. Phys. C {\bf 62}, 455 (1994). 

\bibitem{CMD2opp} CMD-2 Collaboration, R.R.~Akhmetshin {\em et al.},
Phys. Lett. B {\bf 489}, 125 (2000).

\bibitem{CrBar} Crystal Barrel Collaboration, A.V.~Anisovich {\em et al.},
Phys. Lett. B {\bf 485}, 341 (2000). 

\bibitem{E852} E852 Collaboration, P.~Eugenio {\em et al.}, 
Phys. Lett. B {\bf 497}, 190 (2001).

\bibitem{Isgur} S.~Godfrey and N.~Isgur,
Phys. Rev. D {\bf 32}, 189 (1985). 

\bibitem{Dalitz} MARK-III Collaboration,
L.~Chen and W.M.~Dunwoodie, 
in {\em Proceedings of Hadron 91 Conference,
College Park, MD, 1991}, p. 100, SLAC-PUB-5674 (1991).

\end{thebibliography}
\end{document}